\documentclass[11pt,a4paper,twocolumn]{article}
\usepackage{authblk}
\usepackage{titlesec}
\usepackage{lipsum} % For generating dummy text, you can remove this

\usepackage[margin=2cm]{geometry}

\usepackage[utf8]{inputenc}
\usepackage[T1]{fontenc}

\usepackage{graphicx}

\usepackage{xcolor}

\usepackage{authblk}
\usepackage{mathtools, cuted}

\usepackage{bm}
\usepackage{amsfonts}
\usepackage{amsmath}
\usepackage{amssymb}
\usepackage{mathrsfs} % cursive letters
\usepackage{siunitx}
\titleformat{\section}{\centering\bfseries}{\Roman{section}.}{1em}{\MakeUppercase}
\titleformat{\subsection}{\centering\bfseries}{\Alph{subsection}}{1em}{}
\titleformat{\subsubsection}{\centering\itshape}{\arabic{subsubsection}}{1em}{}
\title{\centering\bfseries {Thermal effects on the lifetime of evaporating drops on fibers}}
\author[1,2]{Marie Corpart}
\author[1]{Frédéric Restagno}
\author[1]{François Boulogne}
\affil[1]{Université Paris-Saclay, CNRS, Laboratoire de Physique des Solides, 91405, Orsay, France.}
\affil[2]{Corresponding author: {marie.corpart@gmail.com}}
\date{} % Remove date
\begin{document}

\twocolumn[
    \begin{@twocolumnfalse}
        \maketitle
        \begin{abstract}
        We present an experimental study on the evaporation of drops on fibers. More specifically, we focus on the droplet lifetime both in quiescent air and in an air flow of constant velocity. 
We propose a model to describe the evaporation rate and lifetime in a purely diffusive regime, which includes the liquid cooling associated with evaporation and the thermal conductivity of the atmosphere and the fiber. 
Our model effectively captures the primary physical behaviors, demonstrating a semi-quantiative agreement with our measurements across various liquids and fiber materials. Finally, the model is generalized to a convective air flow, which also rationalizes our experimental data.
        \end{abstract}
    \end{@twocolumnfalse}
]

\section{Introduction}

The simplest consideration for drop evaporation is likely the suspended spherical drop, which has been solved by Langmuir for quiescent air \cite{Langmuir1918} and extended by Frossling for the case of an air flow \cite{Frossling1938}.
Evaporation is associated with a cooling effect due to the enthalpy of vaporization.
The  role of the cooling effect on the droplet lifetime has been studied, for example, by Andrea to understand the evolution of sea spray \cite{Andreas1995}. It can also be used for measuring the relative humidity of air with a psychrometer \cite{Corpart2024}.
A comprehensive model regarding spherical drop evaporation has been developed by Sobac et al.~\cite{Sobac2015}.
Recently, these questions regained interest in the context of the transport of airborne contaminants \cite{Netz2020,Netz2020a}.
Besides evaporation of spherical drops, evaporation of drops deposited on a flat substrate is also an important topic in several applications~\cite{Erbil2012}, including inkjet printing, micro/nano-fabrication, spray cooling and biochemical analysis. However, understanding it in-depth can be challenging due to the complex interplay of several physical phenomena, such as substrate wetting, mass and heat exchanges between the liquid and its environment, and flows within the liquid. Evaporation of sessile drops has been the subjects of several reviews~\cite{Erbil2012, Cazabat2010, Larson2014, Brutin2015,Wilson2023,Brutin2022}.
As demonstrated by numerous studies, the wetting properties of the substrate have a non-negligible effect on the lifetime of a sessile drop by constraining the geometry of the liquid and its temporal evolution~\cite{Picknett1977,Birdi1993, Shanahan1994,Bourges-monnier1995,Rowan1995,Mchale1998,Erbil2002,Soolaman2005,Stauber2014, Stauber2015}.
In the case of a drop deposited on a flat substrate, the cooling of the liquid also has an effect on the evaporation rate. However, the magnitude of the cooling depends on the heat flux exchanged between the drop and the solid, and therefore on the thermal properties of the substrate and the geometry of the system~\cite{David2007, Dunn2009, Sefiane2011, Lopes2013,Larson2014,Schofield2021}. Experimental studies have demonstrated that the liquid cooling significantly slows down liquid evaporation for poorly conducting substrates and/or liquids that evaporate quickly~\cite{David2007, Dunn2009, Sefiane2011, Lopes2013}.
For conductive substrates and low to medium-volatility liquids, drop evaporation can typically be considered quasi-adiabatic~\cite{David2007, Dunn2009, Sefiane2011, Lopes2013}. These experimental results were rationalised by Sefiane and Bennacer~\cite{Sefiane2011}, who employed thermal resistances to model the system.

The presence of the substrate causes the temperature along the interface to be non-uniform~\cite{Dehaeck2014}, resulting in surface tension gradients that can create Marangoni flows~\cite{Hu2005, Ristenpart2007, Larson2014, Xu2010}. 
These flows might affect the rate of evaporation of the droplet~\cite{Girard2008, Semenov2010, Yang2022}.

Although the evaporation of spherical drops and sessile drops on flat substrates have been extensively studied over the past fifty years, research on evaporation in other geometries is scarce. Nevertheless, droplets on fibrous materials also present a relevant situation for many applications, such as drying filters, textiles, and insulation materials~\cite{Pan2006,Sutter2010, Duprat2022}.
The morphology and the evaporation of liquids in fibrous materials are also subject to the influence of the fibres present. For a drop on a single fiber, the liquid can adopt two different morphologies depending on its volume and contact angle. It can take either the clamshell shape, where a small droplet wets only a portion of the fiber's circumference, or the barrel shape, where the droplet fully wets the fiber, resembling a pearl on a string~{\cite{Carroll1976,Chou2011, Mchale2002}}. For the latter, the fiber's curvature creates an inflection point at the liquid/air interface, leading to a droplet aspect ratio close to unity, even for a perfectly wetting liquid, in contrast to what is observed for sessile drops~{\cite{Carroll1976, Brochard1991}}.
In a previous study, we have shown that an axisymmetric drop on a fiber has an evaporation rate that is nearly independent of the contact angle, in contrast to sessile drops \cite{Corpart2022}.
In addition, on an assembly of fibers, the evaporation rate of the liquid depends on the fiber orientation, which mainly changes the liquid morphology \cite{Duprat2013,Boulogne2015a}.

The lifespan of a droplet, whether located at the tip or the center of a single fiber or at the junction of two perpendicular crossed fibers, also depends on the heat exchange between the fiber(s) and the droplet.
This has been verified through numerical simulations, which consider the droplet to be spherical and evaporating in most cases at high pressure and temperature~\cite{Chauveau2019, George2017, Ghata2014, Shih1995,Shringi2013, Yang2001, Yang2002}.
To our knowledge, only one study has been conducted to confirm this experimentally in ambient conditions.
Radhakrishnan et al.~\cite{Radhakrishnan2019} studied the evaporation of ethanol and water drops suspended from the end of a glass or steel fiber at ambient temperature and pressure. They showed that drops evaporate faster on a conductive substrate than on an insulating one.
The above-mentioned studies have also shown that the presence of a fiber can increase the evaporation rate of a drop for a sufficiently large fiber radius.
This is due to the conductive heat flux provided by the fiber to the drop, even for insulating fibers~\cite{Chauveau2019, George2017, Ghata2014, Shringi2013, Yang2001, Yang2002}.
However, this effect decreases with increasing temperature and Reynolds number~\cite{Shih1995, Shringi2013, Yang2001}.
By reducing the cylinder diameter while keeping the drop radius constant, the increase in evaporation flux can be limited or even suppressed.
In this case, the evaporation rate can be accurately estimated using that of a spherical drop~\cite{Chauveau2019, George2017, Yang2001}, eliminating the differences between insulating and conductive substrates. These results are consistent with the analytical model developed by Fuchs for a spherical drop suspended at the end of a fiber, evaporating non-adiabatically in diffusive regime~\cite{Fuchs1959}.
In the case of forced-convective evaporation, a laminar air flow perpendicular to a horizontal fiber leads to drop propulsion along the fiber because of symmetry breaking in the wake behind the drop~\cite{Bintein2019}.
For droplets deposited on an assembly of fibers, air flow can induce aerodynamic interactions between droplets that lead to complex behaviors such as alignment, repulsion, or coalescence~\cite{Wilson2023a}.
Despite the importance of understanding the evaporation of drops in fibers, there have been few experimental studies or attempts to rationalize the results.

Here, we aim to study experimentally and theoretically the evaporation of a single axisymmetric drop on a fiber both in quiescent air and under an air flow.
We will  highlight the role of the fiber thermal properties on the drop lifetime.
In Section~\ref{sec:experiment}, we present the experimental setup and the measurements of the droplet lifetime on fibers. In Section~\ref{sec:cooling}, we propose an analytical model of the drop evaporation including the evaporation-induced cooling effect in a diffusion-limited regime that we compare to the experimental results in absence of air flow.
In Section~\ref{sec:air-flow}, we extend the model to include the air flow to rationalize our entire dataset.
Then, we conclude in Section~\ref{sec:conclusion}.

%%%%%%%%%%%%%%%%%%%%%
%
%%%%%%%%%%%%%%%%%%%%%
\section{Experiments}\label{sec:experiment}
\begin{figure}
    \centering
    \includegraphics[width = 1\linewidth]{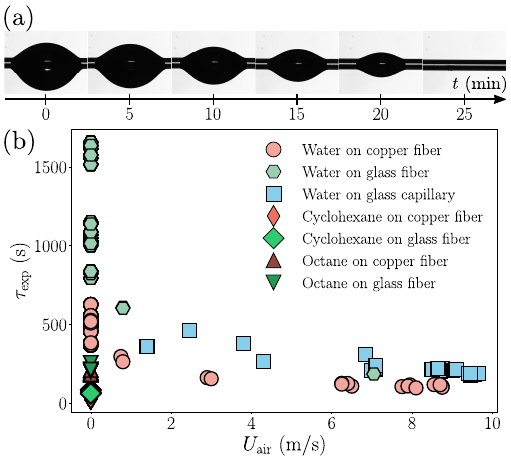}
    \caption{(a) Time-lapse of the evaporation of a water drop of initial volume $\Omega_0 \approx 0.8$~µL deposited on a glass fiber of radius $a = 125$~µm observed in side view. The measured relative humidity is ${\cal R}_H^{\rm exp} = 45.9~\%$ and the temperature is $T_{\rm exp} = 21.5~^\circ$C. (b) Measured lifetime of water, cyclohexane and octane droplets on copper and glass fibers ($\Omega_0 = 0.8~\rm \mu$L, $a = 125~\rm \mu$m.) as a function of the air velocity $U_{\rm air}$.}
    \label{fig:timelapse_and_raw_data}
\end{figure}
\subsection{Drops on fibers} To measure experimentally the lifetime $\tau_{\rm exp}$ of a drop on a fiber, we
used fibers of diameter $2a = 250$~µm which are either glass fibers supplied by Saint-Gobain, glass capillaries (inner diameter 150~µm, VitroCom \#CV1525), or copper fibers (Goodfellow CU005270). The surface roughness of the fibers has not been measured but is below $1 ~{\mu}$m as can be deduced from optical imaging of the fibers.
A drop of volume $0.8$~µL is deposited on the fiber with a pipette (Eppendorf 0.1 -- 2.5~µL).
Before the deposition, the surface of the fiber is activated with a plasma generator (Electro-Technics Product).
The studied liquids are distilled water (resistivity: 18.2~M$\Omega\cdot$cm), cyclohexane (Sigma, purity $> 99.5~\%$) and octane (Sigma, purity $> 99.5~\%$). For the chosen volume and fiber diameter the droplet can adopt either the clamshell or the barrel conformation (metastable region of the morphology diagram~\cite{Chou2011}). 
Lifetimes of droplets are measured only for drop adopting an axisymmetric morphology (barrel shape, see Figure~\ref{fig:timelapse_and_raw_data}(a) and Figure~A.2 of the Supplementary Materials) on the fiber~{\cite{Carroll1976, Chou2011, Mchale2002}}. In the range of volume and contact angle where the barrel morphology is stable, the macroscopic shape of the drop is relatively independent of wetting conditions, as illustrated in Figure A.2 of the Supplementary Materials.
\subsection{Wind tunnel}
Experiments are performed in a homemade wind tunnel either in diffusive regime (still air) or in forced convective regime. A photograph and a schematic representation of the setup are provided in Figure A.1 of the Supplementary Materials.
The presence of the box limits the sensitivity of the system to air movements in the room.
For forced-convective experiments, the air is set in motion by a fan (Soler \& Palau, TD 6000/400, 5310 m$^3$/h 230~V) and is aspirated into a squared-section box of dimensions $110 \times 40 \times 40$~cm$^3$, opened at the entrance.
The fiber is placed horizontally in the center of the box perpendicularly to the air flow.
The air flow velocity is measured with a hot wire anemometer (Radiospare AM-4204 RS PRO) allowing to measure velocities ranging from 0.2 to 20~m/s with a resolution of 0.1~m/s.
The fan is connected to a variable transformer allowing to impose a voltage between 0 and 260~V (CONATEX Variable transformer 0 to 260V/3A AC) corresponding to air velocities $U_\textrm{air}$ between 0.1 and 10~m/s.
Denoting $\nu_{\rm air} \approx 1.5 \times 10^{-5}~\rm m^2/s$ as the kinematic viscosity of air, the associated Reynolds numbers, defined as $\mathrm{Re} = 2 R U_{\rm air}/\nu_{\rm air}$, are hence between 10 and 1000 for a typical drop size of $2R=1$~mm.

\subsection{Observations and lifetime measurements}

The observation is done in lateral view with a homemade microscope equipped with a 2$\times$ long working distance objective (Mitutoyo) connected to a monochrome camera (Basler acA5472-17um, resolution: 20 Mpx). Images are acquired at 1 fps.
Temperature and relative humidity can vary in the room where the experiments are performed and are measured with a commercial hygrometer for each experiment. These two parameters are denoted $T_{\rm exp}$ and $\mathcal{R}_\mathrm{H}^{\rm exp}$, respectively. 
A time lapse of a water drop evaporating on a glass fiber is shown in Figure~\ref{fig:timelapse_and_raw_data}(a). Lifetimes $\tau_\textrm{exp}$ are measured by recording the time between the first image where we see the drop and the first image where we see no liquid on the fiber.
Figure B.1 of the Supplementary Materials depicts the lifetimes of droplets, measured in the diffusive regime, as a function of relative humidity and temperature. The typical measurement error is approximately ten seconds per point, and the error bars are smaller than the markers. 
In terms of measurement repeatability, for a specific liquid-solid system at a fixed temperature and humidity, the deviation of the lifetime measurements on multiple drops is approximately 10 \%. These discrepancies are primarily attributable to the stages involved in the pipetting of the drop and its deposition on the fiber, both of which influence the initial volume of liquid. Consequently, we endeavor to replicate measurements under identical experimental conditions whenever feasible.

For forced convective experiments, the deposition of the drop on the fiber is done at $U_\textrm{air} =0$, and then once the focusing on the drop is done, the wind tunnel is turned on.
Thus, we estimate that the uncertainty of the lifetime measurement is about 30~s. 
As a result, the relative uncertainties on the high speed lifetime measurements ($\tau_\textrm{exp} \approx 100$~--~200~s) are relatively large.

Figure~\ref{fig:timelapse_and_raw_data}(b) depicts the measured lifetimes as a function of air velocity $U_{\rm air}$ for the various tested liquids and fibers.
The measurements appear scattered, and Figure~\ref{fig:timelapse_and_raw_data}(b) shows that the lifetimes decrease with air velocity. To understand these results we first rationalize the observations for the diffusion limited evaporation regime at $U_{\rm air} = 0$. Then, we will consider the effect of air velocity on the lifetime of the drop.

%%%%%%%%%%%%%%%%%%%%%
%
%%%%%%%%%%%%%%%%%%%%%
\section{Evaporative cooling in the diffusion-limited regime}\label{sec:cooling}
\subsection{Models}
\subsubsection{Spherical drop}
In a previous study~\cite{Corpart2022} we showed that the evaporation rate of a barrel-shaped drop on a fiber is virtually independent of contact angle and can be approximated by the one of a spherical drop evaporating in the same conditions.
Therefore, we consider a spherical airborne droplet evaporating in purely  diffusive regime, $U_{\rm air} = 0$.

For a spherical drop of radius $R(t)$ evaporating in still air at temperature $T_\infty$ and relative humidity $\mathcal{R}_\mathrm{H}$, the diffusion-limited evaporation is in good approximation quasi-stationary and the evaporation rate writes~\cite{Fuchs1959,Sobac2015}
\begin{equation}\label{eq:sphere_Phi_ev_diff}
    \Phi_{\rm ev} = 4\pi \mathcal{D} R \left(c_{\rm sat}( T_{\rm i}) - c_\infty\right).
\end{equation}
The quasi-steady assumption is valid if the droplet radius is significantly larger than the mean-free path of the vapor molecules, meaning $R$ larger than few micrometers~{\cite{Fuchs1959}} and if the evaporation time of the drop is much larger than the mass diffusion characteristic time in the gas phase $R^2/\cal D$~{\cite{Sobac2015}}. In practice, we check that this is the case.
The enthalpy of vaporization $\Delta_\text{vap}H$ being a positive quantity, the liquid cools down while evaporating and the interface reaches a temperature $T_{\rm i} \leq T_\infty$.

In equation~(\ref{eq:sphere_Phi_ev_diff}), $\mathcal{D}$ is the diffusion coefficient of water vapor in air at the temperature of the interface.
In a previous study, we have shown that the value of the diffusion coefficient can be estimated at the temperature of the air far from the drop $T_\infty$~\cite{Corpart2023}.
At the interface the air is saturated with vapor, $c(r = R) = c_{\rm sat}(T_{\rm i})$ where $c_{\rm sat}(T_{\rm i})$ is the saturated vapor concentration at the temperature of the interface.
Far from the interface, the vapor concentration is the ambient concentration noted $c_\infty$, which -- in the hypothesis that the vapor is an ideal gas -- can be related to the relative humidity by $\mathcal{R}_\mathrm{H} = c_\infty/c_{\rm sat}(T_\infty)$ where $c_{\rm sat}(T_\infty)$ is the saturated vapor concentration at the temperature of the ambient air.
Integrating the mass conservation $\Phi_{\rm ev} = - 4/3 \pi \rho\, \mathrm{d}R^3/\mathrm{d}t$ associated with equation~(\ref{eq:sphere_Phi_ev_diff}), the lifetime of a spherical drop evaporating in diffusive regime is obtained and reads
\begin{equation}\label{eq:sphere_tau_ev_diff}
    \tau_{\rm S} = \frac{\rho R_0^2}{2 \mathcal{D} \left(c_{\rm sat}(T_{\rm i}) - c_\infty \right)},
\end{equation}
where $R_0$ is the initial radius of the drop and $\rho$ the density of the liquid.
To calculate the lifetime $ \tau_{\rm S}$, the temperature of the liquid $T_{\rm i}$ must then be determined.

Due to the cooling of the liquid the drop exchanges heat with its environment. By analogy with mass transfer, the total heat flux $Q_{\rm h}$ received by the spherical drop from the air in purely diffusive and quasi-stationary regime is 
\begin{equation}\label{eq:sphere_Q_h_diff}
    Q_{\rm h} = - 4\pi \lambda_{\rm air} R \Delta T^\star,
\end{equation}
with $\lambda_\textrm{air}$ the thermal conductivity of the air at $T_\infty$ and $\mathcal{R}_\mathrm{H} = 0$~\cite{Corpart2023}.
We note $\Delta T^\star = T_\infty - T_{\rm i}$, to get more compact equations. 
The steady-state assumption implies that the heat diffusion in the air and in the liquid are fast compared to the lifetime of the drop. 
This is validated if the timescales over which the heat diffuses through the air $R_0^2/\alpha_{\rm air}$ and through the liquid $R_0^2/\alpha_{\ell}$  with $\alpha_{\rm air }$ and $\alpha_{\ell}$ the thermal diffusivities of the air and the liquid, are short compared to the evaporative time. This also implies that the temperature in the drop is uniform and has reached its equilibrium value $T_{\rm i}$~{\cite{Fuchs1959,Sobac2015}}. 
In practice, these conditions are valid for the tested liquids evaporating under ambient conditions~{\cite{Beard1971, Andreas1995, Sobac2015, Netz2020, Netz2020a}}.

In the steady state regime, the energy absorbed by the evaporation $Q_{\rm ev} =  \Delta_\text{vap}H \, \Phi_{\rm ev}$ balances the thermal flux such as
\begin{equation}\label{eq:sphere_balance_energy_diff}
    Q_{\rm ev} = - Q_{\rm h},
\end{equation}

\begin{equation}\label{eq:delta_T_sphere_diff_implicite}
   T_\infty - T_{\rm i} = \chi\left( \frac{c_{\rm sat}(T_{\rm i})}{c_{\rm sat}(T_\infty)} - \mathcal{R}_{\rm H}\right).
\end{equation}
To get an analytical prediction of $\Delta T^\star$, the evolution of the saturating vapor concentration with temperature is approximated by~\cite{Corpart2023}:
\begin{equation}\label{eq:quadratic_c_sat_vs_T}
    \frac{c_\textrm{sat}(T)}{  c_\textrm{sat}(T_\infty)} = 1 + \alpha_1  \Delta T + \alpha_2  \Delta T^2,
\end{equation}
where $\Delta T = T_\infty -T$ and $\alpha_1$ and $\alpha_2$ are two fitting parameters depending only on $T_\infty$. 
The values of $\alpha_1$ and $\alpha_2$ are given in Table~4 of the Supplementary Materials for $T_\infty \in [10, 30]~^\circ {\rm C}$ for all the studied liquids.
Inserting equation~(\ref{eq:quadratic_c_sat_vs_T}) into equation~(\ref{eq:sphere_balance_energy_diff}) (see~\cite{Corpart2023} or Supplementary Materials~C.1 for detailed resolution), we obtain the liquid temperature:
\begin{equation}\label{eq:sphere_Delta_T}
    \Delta T^\star = \frac{ 1 -  \chi \alpha_1  - \sqrt{\left(1 - \chi \alpha_1 \right)^2 - 4 \chi^2 \alpha_2 \left(1 - \mathcal{R}_{\rm H} \right)}}{2 \chi \alpha_2},
\end{equation}
where $\chi = \Delta_{\rm vap} H {\cal D} c_{\rm sat}(T_\infty)/  \lambda_{\rm air}$ depends only on $T_\infty$.
The physical parameters $c_{\rm sat}$, $\mathcal{D}$ and $\lambda_{\rm air}$ are evaluated at $T_\infty$ by using the phenomenological equations introduced in~\cite{Corpart2023} and recalled in the Supplementary Materials~B.
The liquid density $\rho$ and enthalpy of vaporization $\Delta_\text{vap}H$ are considered constant and their values are given for all the studied liquids in the Table~1 of Supplementary Materials.

Finally, combining equations~(\ref{eq:quadratic_c_sat_vs_T}) and (\ref{eq:sphere_Delta_T}), and inserting the result in equation~(\ref{eq:sphere_tau_ev_diff}) gives the lifetime of the spherical drop.

\subsubsection{Axisymmetric drop on a fiber}
\begin{figure}[t]
    \centering
    \includegraphics[width = 1\linewidth]{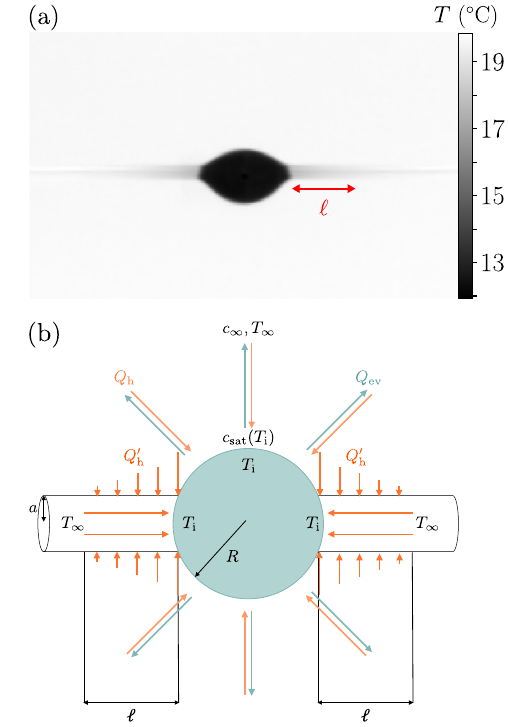}
    \caption{(a) Image captured with an infrared camera (Teledyne FLIR X6981 with a 50~mm macro objective) of a 0.8 $\mu$L drop on a glass fiber.
    The gray scale encodes the temperature.
    The image highlights the temperature gradient along the fiber on both sides of the drop.
    (b) Schematic representation of the heat fluxes exchanged between the drop on a fiber and its environment with the notations used for the modeling.
    }
    \label{fig:schema_heat_fluxes}
\end{figure}

% \TODO{Insister sur l'independance de Qev avec l'angle de contac t--> pas besoin de connaitre le mode d'evap pour predire le temps de vie ET resultat de notre papier manchon que Qev = Qsphere }
Due to the different surface tensions of the three liquids, their wettability on the wires vary.
However, these variations only cause minor changes of the droplet shapes, as expected from the literature~{\cite{Carroll1976,Chou2011, Mchale2002}} and as shown in the time series presented in the supplementary information (Figure A.2).  
Furthermore, a previous study demonstrated that the contact angle has a weak effect on the evaporation rate and that axisymetric drops on a fiber evaporate, in a good approximation, as spherical droplets~{\cite{Corpart2022}}.
To model the diffusion-limited evaporation of an axisymmetric drop on a fiber, we thus adapt the model developed by Fuchs~\cite{Fuchs1959} for a spherical drop suspended at the end of a fiber and we consider the system shown in Figure~\ref{fig:schema_heat_fluxes}(b) of a spherical drop of radius $R$ pierced by a fiber of radius $a$. 
The drop evaporates in still air which is at temperature $T_\infty$ and relative humidity $\mathcal{R}_\mathrm{H}$.
The drop exchanges mass and heat with its environment and the goal here is to calculate the heat flux $Q'_{\rm h}$ exchanged between the fiber and the liquid on each side of the drop (see Figure~\ref{fig:schema_heat_fluxes}(b)).
The calculation of $Q'_{\rm h}$ takes into account heat conduction through the fiber and the heat flux exchanged between the fiber and the atmosphere.
The length over which the temperature gradient in the fiber is established is denoted $\ell$.
In the steady state regime, the energy absorbed by the evaporation balances the thermal flux such as
\begin{equation}\label{eq:fiber_balance_energy_diff}
    Q_{\rm ev} = - (Q_{\rm h} + 2Q'_{\rm h}).
\end{equation}

To perform the calculations, we make additional hypothesis. 
First, we consider systems where gravity is negligible meaning that $\mathrm{Wo} = \rho g \Omega/(\gamma a) \ll 1$ with $\mathrm{Wo}$ the Worthington number~\cite{Worthington1885} where $\Omega$ is the volume of the drop and $\gamma$ the surface tension.
We also assume that the exchanges between the drop and the atmosphere are not perturbed by the presence of the fiber and are those of a spherical droplet (see Figure~\ref{fig:schema_heat_fluxes}(b)).
The energy absorbed by the evaporation is thus $Q_{\rm ev} =  \Delta_\text{vap}H \, \Phi_{\rm ev}$ with $\Phi_{\rm ev}$ defined in equation~(\ref{eq:sphere_Phi_ev_diff}) and the heat flux exchanged between the atmosphere and the drop is $Q_{\rm h}$ defined in equation~(\ref{eq:sphere_Q_h_diff}).
Finally, the temperature of the drop is assumed to be uniform both in the volume and on the surface, the latter being supported by the infrared image shown in Figure~\ref{fig:schema_heat_fluxes}(a), such that the temperature of the drop is the temperature of the interface $T_{\rm i}$.
The temperature inside the fiber is also supposed to be constant in the cross section of the fiber $T(z)$.
This assumption requires that $ a \ll \ell$, which is also supported by the Figure~\ref{fig:schema_heat_fluxes}(a).

To calculate the temperature profile $T(z)$ in the fiber, we consider only one of the two sides of the fiber and adopt a cylindrical coordinate system centered on the liquid/solid interface (see Figure~D.1(b) of the Supplementary Materials).
The heat flux $Q'_{\rm h}$ is defined as
\begin{equation}\label{eq:Q_prime_h_def}
    Q{'}_h = - \pi a^2 \lambda_\textrm{s} \left. \frac{\mathrm{d}T}{\mathrm{d}z} \right|_{z = 0},
\end{equation}
where $\lambda_\textrm{s}$ is the thermal conductivity of the fiber.
In addition, the local energy balance writes:
\begin{equation}\label{eq:local_energy_balance_in_the_fiber}
    -2 \pi a \, \lambda_\textrm{air} \,  \left. \frac{\mathrm{d}T}{\mathrm{d}r} \right|_{r = a} = \pi a^2 \lambda_{\rm s}  \frac{\mathrm{d}^2T}{\mathrm{d}z^2}.
\end{equation}
The resolution of this differential equation is detailed in the Supplementary Materials~D and leads to the estimation of the total heat flux exchanged between the fiber and the drop:
\begin{equation}\label{eq:Q_prime_h}
    Q{'}_\textrm{h}  \approx - \pi a \Delta T^\star   \sqrt{\frac{\pi \lambda_\textrm{air} \lambda_\textrm{s}}{10}}.
\end{equation}
Substituting equation~(\ref{eq:Q_prime_h}) in equation~(\ref{eq:fiber_balance_energy_diff}), we get an implicit equation for the liquid temperature $T_{\rm i}$:
\begin{equation}\label{eq:Delta_T_vs_Delta_c_sphere_on_fiber}\begin{split}
    T_\infty - T_{\rm i} \left(1 + \Tilde{\mathcal{Q}}_{\rm fiber} \right) = {\chi}\left( \frac{c_{\rm sat}(T_{\rm i})}{c_{\rm sat}(T_\infty)} - \mathcal{R}_{\rm H}\right),
\end{split}
    \end{equation}
    where
    \begin{equation}\label{eq:Tilde_Q_fibre}
        \Tilde{\mathcal{Q}}_{\rm fiber} = \frac{a}{R}\sqrt{\frac{\pi \lambda_\textrm{s}}{40 \lambda_{\rm air}}}
    \end{equation}
    is the dimensionless number computing the ratio between the heat fluxes exchanged between the air and the liquid and between the solid and the liquid.
    Using the quadratic description of equation~(\ref{eq:quadratic_c_sat_vs_T}) together with the method proposed in Ref.~\cite{Corpart2023}, we obtain the temperature of the drop from equation~(\ref{eq:Delta_T_vs_Delta_c_sphere_on_fiber}):

    \begin{equation}\label{eq:Delta_T_sphere_on_fiber}\begin{split}
        \Delta T^\star = &\frac{ 1 + \Tilde{\mathcal{Q}}_{\rm fiber} -   \chi \alpha_1}{2  \chi \alpha_2} - \\ & \frac{\sqrt{\left[1 +  \Tilde{\mathcal{Q}}_{\rm fiber} -  \chi \alpha_1 \right]^2 - 4  \chi^2 \alpha_2 \left(1 - \mathcal{R}_{\rm H} \right)}}{2  \chi \alpha_2},
    \end{split}
\end{equation}
which depends on the drop radius $R$, contrarily to the case of a isolated spherical drop, \textit{via} $\Tilde{\mathcal{Q}}_{\rm fiber}$.

\subsection{Results and Discussion}

\begin{figure}[h!]
    \centering
    \includegraphics[width = 1\linewidth]{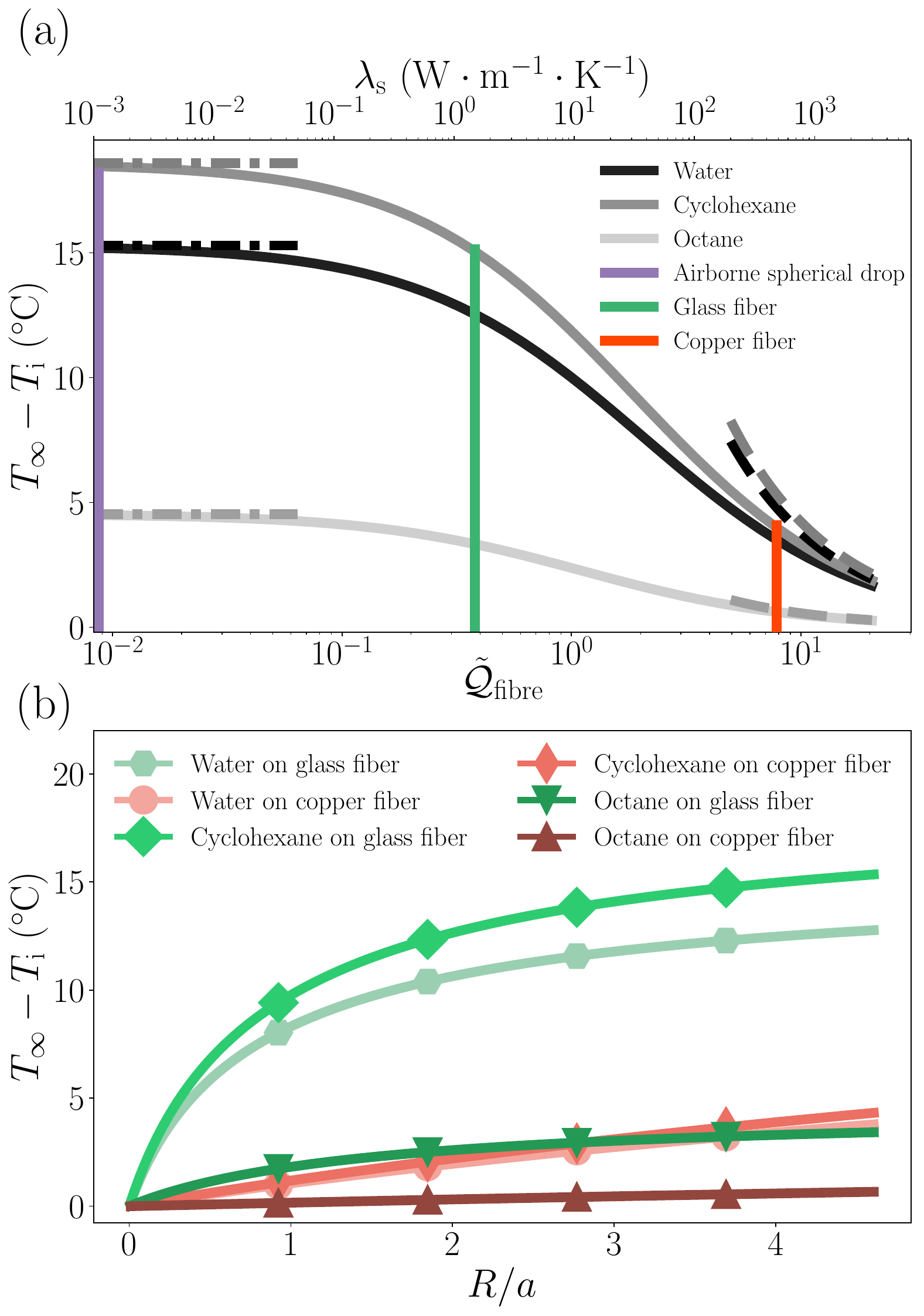}
    \caption{
        Prediction of the temperature difference $\Delta T^\star$ for a spherical drop on a fiber (Eq.~(\ref{eq:Delta_T_sphere_on_fiber})) for $\mathcal{R}_{\rm H } = 0$ et $T_\infty = 20~^\circ \rm C$.
        (a) Effect of the thermal properties of the fiber $\lambda_{\rm s}$ on liquid temperature for $a = 125~{\rm \mu m}$ and $R = R_0 \approx 0.6~{\rm mm}$.
        The curves in gray shades are for the three different liquids and the vertical lines indicates three different materials. In purple we show a line for $\Tilde{\mathcal{Q}}_{\rm fiber} \to 0$, which corresponds to the limit case of a isolated airborne sphere (see  Eqs.~({\ref{eq:delta_T_sphere_diff_implicite}}) and ({\ref{eq:Delta_T_vs_Delta_c_sphere_on_fiber}})).  
        The dashed-dotted lines represent the limit of equation~({\ref{eq:Delta_T_sphere_on_fiber}}) for $\tilde{\cal Q}_{\rm fiber} \ll 1 - \chi\alpha_1 \sim 1$ which corresond to the case of an airborne sphere (see Eq.~31 of the Supplementary Materials). The dashed lines represents the limit of equation~({\ref{eq:Delta_T_sphere_on_fiber}}) for $\tilde{\cal Q}_{\rm fiber} \gg  1 - \chi\alpha_1 \sim 1$ for each liquid (see equation~(32) of the Supplementary Materials).
        (b) Effect of the aspect ratio of the drop $R/a$ for the three liquids and two fiber materials.
        The markers are visual guides to distinguish between the different curves.}
    \label{fig:delta_T_vs_lambda_s_et_R}
\end{figure}
The dimensionless number $\tilde{{\cal Q}}_{\rm fiber}$ (Eq.~(\ref{eq:Tilde_Q_fibre})) indicates the importance of the fiber contribution to the heat exchange by taking into account the geometry of the system \textit{via} $a/R$ and the difference in thermal conductivity between the air and the fiber.

If $\Tilde{\mathcal{Q}}_{\rm fiber} \ll 1$, then equations~(\ref{eq:Delta_T_vs_Delta_c_sphere_on_fiber}) and~({\ref{eq:delta_T_sphere_diff_implicite}}) show that the effect of the fiber on the heat exchange is negligible.
Therefore, the temperature of the liquid depends only on its physical properties, as if it were evaporating as an isolated sphere.
This scenario is observed for highly insulating materials and/or a small fiber relative to drop size.

For $\Tilde{\mathcal{Q}}_{\rm fiber} \gg 1$, the heat flux brought by the fiber to the drop is large compared to the one brought by the atmosphere: the fiber thermalizes the drop at each moment, the temperature of the liquid tends towards the ambient temperature meaning that the evaporation of the liquid is done in a quasi-adiabatic way.
This situation is observed for conductive fibers.
The case where $a \gg R$ is also possible, but it leads to an equilibrium morphology of \textit{clamshell}~{\cite{Carroll1976, Chou2011, Mchale2002}}, which is not addressed in this work.

To analyze the effect of liquid physicochemical properties, fiber thermal conductivity, and system geometry on liquid temperature, the temperature difference between liquid and ambient air is plotted as a function of the fiber thermal conductivity (Figure~\ref{fig:delta_T_vs_lambda_s_et_R}(a)) or the drop radius (Figure~\ref{fig:delta_T_vs_lambda_s_et_R}(b)).
The curves are obtained from equation~(\ref{eq:Delta_T_sphere_on_fiber}) at $T_\infty = 20~^\circ {\rm C}$ and $\mathcal{R}_{\rm H} = 0$.

Figure~\ref{fig:delta_T_vs_lambda_s_et_R}(a) illustrates the variation of the liquid temperature with respect to the thermal conductivity of the fiber, $\lambda_{\rm s}$ or $\Tilde{\mathcal{Q}}_{\rm fiber}$, for a dimensionless drop radius $R_0/a \approx 5$, which corresponds to the initial dimensions of the experimental system. 
The thermal conductivity of the solid, $\lambda_{\rm s}$, varies from $1 \times 10^{-3}~\mathrm{W \cdot m^{-1} \cdot K^{-1}}$ (one-tenth of the air conductivity) to $1000~\mathrm{W \cdot m^{-1} \cdot K^{-1}}$, which would correspond to a fiber made of diamond. 
For instance, the thermal conductivity of glass is about $1~\mathrm{W \cdot m^{-1} \cdot K^{-1}}$. and copper is $400~\mathrm{W \cdot m^{-1} \cdot K^{-1}}$. 
Thus, the prefactor $\sqrt{ \pi \lambda_\textrm{s} / 40 \lambda_{\rm air}}$ in equation ({\ref{eq:Tilde_Q_fibre}}) is 1.6 and 32 for glass and copper, respectively.
% \marie{Pourquoi ne pas donner la valeur de Qfibre (0.3 et 7) ?}
The different liquids under study are depicted in shades of gray.
The violet curve illustrates the case of an airborne spherical drop, $\Tilde{\mathcal{Q}}_{\rm fiber} \to 0$, and its intersection with the curves representing the various liquids provides the value of temperature, $T_{\rm i}$, obtained from equation~(\ref{eq:Delta_T_sphere_on_fiber}). 
%We also plot in figure~\ref{fig:delta_T_vs_lambda_s_et_R}(a), the limits $\Tilde{\mathcal{Q}}_{\rm fiber} \rightarrow 0$ and $\Tilde{\mathcal{Q}}_{\rm fiber} \rightarrow 1 - \chi \alpha_1$
%\TODO{R3 Description des asymptotes à grand et petit Qfibre de la nouvelle Fig 3b}
Figure~\ref{fig:delta_T_vs_lambda_s_et_R}(a) shows that the effect of evaporation cooling is less pronounced for octane than for cyclohexane and water.
%This finding is supported by Figure~C.1(a) in the Supplementary Materials, where the measured lifetimes of octane drops placed on glass or copper fibers are well described by the model of an adiabatic evaporation of a spherical drop defined in equation~(\ref{eq:sphere_tau_ev_diff}) for $T_{\rm i} = T_\infty$. 

Conversely, Figures~\ref{fig:delta_T_vs_lambda_s_et_R} and C.1(a) in the Supplementary Materials reveal that on a glass fiber, the cooling of water and cyclohexane drops cannot be disregarded.
Furthermore, it can be observed in Figure~\ref{fig:delta_T_vs_lambda_s_et_R}(a) that the temperature within a drop on a glass fiber is higher than that of an airborne spherical drop.
Despite glass being a good insulator, the presence of the fiber cannot be ignored in the thermal exchanges between the drop and its surroundings.
The value of $\Tilde{\mathcal{Q}}_{\rm fiber}$ at the initial instant, i.e., for $R_0/a \approx 5$, $\Tilde{\mathcal{Q}}_{\rm fiber} \approx 0.4 \sim 1$ , aligns well with this observation.
Figure~\ref{fig:delta_T_vs_lambda_s_et_R}(b) indicates that even as $R/a$ decreases over time, the drop's cooling must be considered.
It is only in the final moments of evaporation that the temperature of a drop placed on an insulating fiber equals the ambient temperature.
These observations are experimentally confirmed, as shown in Figure~C.1(b) of Supplementary Materials, which illustrates the lifetime of different tested liquids placed on a glass fiber as a function of the lifetime of an airborne droplet cooling down while evaporating (Eq.~(\ref{eq:sphere_tau_ev_diff}) combined with Eq.~(\ref{eq:sphere_Delta_T})).
It shows that there is a systematic discrepancy between the experimental results for cyclohexane and water and the airborne sphere model. In a broader context, Figure~\ref{fig:delta_T_vs_lambda_s_et_R}(a) shows that for most insulating materials, the presence of the fiber significantly affects thermal exchanges between the drop and its environment and shortens its lifetime.
\begin{figure}
    \centering
    \includegraphics[width = 1\linewidth]{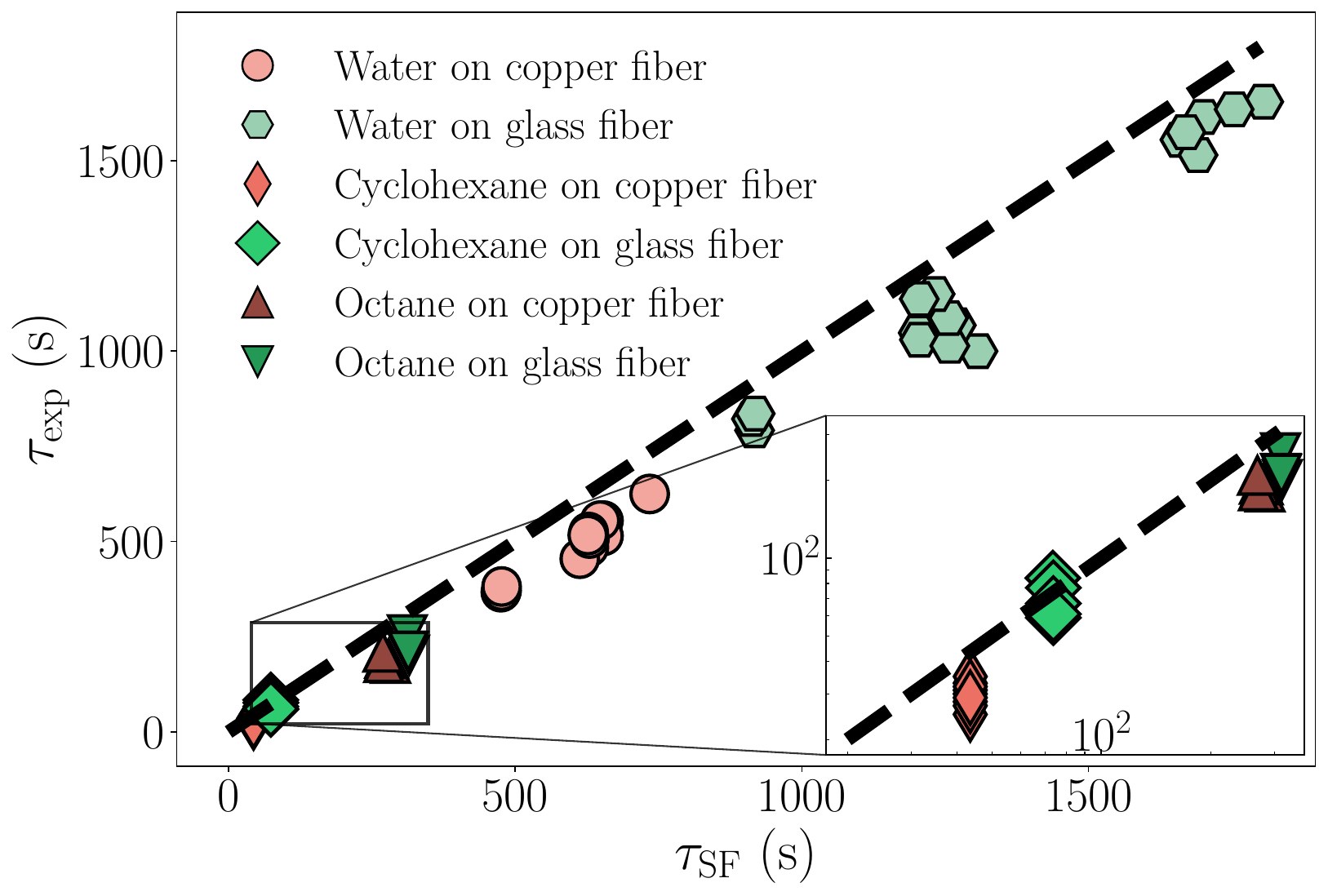}
    \caption{Data presented in Figure~\ref{fig:timelapse_and_raw_data} for $U_{\rm air} = 0$ where the measured lifetime is plotted against the numerical solution of equation \ref{eq:eq_diff_R_de_t_sphere_on_fiber}.
    The black dashed line represents the equality between axes.}
    \label{fig:tau_exp_vs_tau_theo_DS}
\end{figure}

Finally, to predict the lifetime of the drop we substitute equation~(\ref{eq:Delta_T_sphere_on_fiber}) in equation~(\ref{eq:sphere_Phi_ev_diff}), which gives the evaporation rate of the drop.
From the mass conservation of the system, we obtain the following differential equation for the temporal evolution of the drop radius:
\begin{equation}\label{eq:eq_diff_R_de_t_sphere_on_fiber}
    \begin{split}
        - \frac{\rho R \dot{R} }{\mathcal{D} c_\textrm{sat}(T_\infty)}  = \alpha_2 {\Delta T^\star}^{2} + \alpha_1  \Delta T^\star + 1 - \mathcal{R}_{\rm H},
    \end{split}
\end{equation}
where $\Delta T^\star$ is given by equation~(\ref{eq:Delta_T_sphere_on_fiber}).
We integrate numerically the previous equation between $R = R_0$ and $R = 0$ using \textsf{odeint} from scipy \cite{Jones2001} to get the lifetime of the drop on a fiber predicted by the model $\tau_{\rm SF}$.

We plot in Figure~\ref{fig:tau_exp_vs_tau_theo_DS} the measured lifetimes as a function of the theoretical lifetimes $\tau_{\rm SF}$ predicted by the model.
For each point the theoretical lifetime is calculated for the temperature $T_\infty^{\rm exp}$ and humidity ${\cal R}_{\rm H}^{\rm exp}$ as explained in Supplementary Materials~B.
The error bars on each point are smaller than the marker. The repeatability error is given by the height of the cloud of points with a constant $\tau_{\rm SF}$.
The model overestimates (with a difference of the order of 10 -- 20~\%)  the measured lifetimes.
This discrepancy might arise from errors in determining the initial volume, which has a substantial influence as the lifetime depends on $R_0^2$.
Additionally, the non-spherical geometry of the drop, the deformation of the drop by gravity or the impact of natural convection and air movements in the room, which accelerate drying, might contribute to the observed deviations between the model and the experiments.

Now that the case of diffusion-limited evaporation has been explained, let us examine the impact of air velocity on the droplet's drying rate.
%\subsubsection{Lifetime of the drops}

%%%%%%%%%%%%%%%%%%%%%
%
%%%%%%%%%%%%%%%%%%%%%
\section{Role of the air velocity}\label{sec:air-flow}

\subsection{Models}
\subsubsection{Spherical drop}
We consider first the evaporation of a spherical airborne drop of radius $R$ placed in a laminar external air flow of velocity $U_{\rm air}$.
In 1938, Frössling~\cite{Frossling1938}, using boundary layers theory, showed that the evaporation rate of a spherical drop placed in a laminar air flow writes:
\begin{equation}\label{eq:Phi_ev_conv_sphere}
    \Phi_{\rm ev}^{\rm conv} =  \Phi_{\rm ev}\left(1 + \beta_{\rm ev} \textrm{Re}^{1/2} \textrm{Sc}^{1/3}\right),
\end{equation}
where $\Phi_{\rm ev}$ is the evaporation rate of the drop in purely diffusive regime (Eq.~(\ref{eq:sphere_Phi_ev_diff})), $\mathrm{Re} = 2 R U_{\rm air}/\nu_{\rm air}$ is the Reynolds number, $\mathrm{Sc} = \nu_{\rm air}/\mathcal{D}$ the Schmidt number and $\beta_{\rm ev} \approx 0.3$ is a constant.

Analogously, the convective heat flux received by the drop from the atmosphere can be written as follows~\cite{Ranz1952a, Ranz1952}:
\begin{equation}\label{eq:Q_h_conv_sphere}
    Q_{\rm h}^{\rm conv} =  Q_{\rm h}\left(1 + \beta_{\rm h} \textrm{Re}^{1/2} \textrm{Pr}^{1/3}\right),
\end{equation}
where $\mathrm{Pr} = \nu_{\rm air}/{\alpha_{\rm air}}$ is the Prandtl number and $\beta_{\rm h} \approx 0.3$.
The air thermal diffusivity is $\alpha_{\rm air} = \lambda_{\rm air}/(\rho_{\rm air} C_{\rm p}^{\rm air})$ where $\rho_{\rm air}$ is the air density, $C_{\rm p}^{\rm air}$ its specific heat capacity at constant pressure, and $\lambda_{\rm air}$ its thermal conductivity.

In gases, kinetic theory models show that the microscopic mechanism of momentum, mass and thermal diffusion are of the same origin meaning that $\mathrm{Sc} \approx \mathrm{Pr} \approx 1$.
Writing the energy balance in quasi-steady state, $\Delta_\text{vap}H \, \Phi_{\rm ev}^{\rm conv} = - Q_{\rm h}^{\rm conv}$, we get the temperature of the liquid that is virtually independent on air velocity or drop radius and equal to the temperature in purely diffusive regime of equation~(\ref{eq:sphere_Delta_T}) (see Supplementary Materials~C.2).

We denote $x =  \beta_{\rm ev} \textrm{Re}_0^{1/2}\textrm{Sc}^{1/3}$ and $\textrm{Re}_0 = 2R_0 U_{\rm air}/\nu_{\rm air}$ the initial Reynolds number.
Combining equations~(\ref{eq:quadratic_c_sat_vs_T}), (\ref{eq:sphere_Delta_T}), and (\ref{eq:Phi_ev_conv_sphere}) to write mass conservation and integrating the differential equation obtained (see Supplementary Materials~C.2 for detailed resolution) the lifetime of the spherical droplet writes
\begin{equation}\label{eq:SPHERE_tau_CF_full}
    \begin{cases}
        \tau_{\rm S}^{\rm conv}  =  \tau_{\rm S} &\text{ if } x = 0,\\
        \tau_{\rm S}^{\rm conv}  =  \tau_{\rm S}\,{\cal F}(x) &\text{ if } x > 0,
    \end{cases}
\end{equation}
with $\tau_{\rm S}$ the diffusive lifetime of the drop (Eq.~(\ref{eq:sphere_tau_ev_diff})) and 
\begin{equation}
    {\cal F}(x) = \frac{4 }{3 x} - \frac{2}{x^2} + \frac{4 }{x^3} - \frac{4\,\ln{(1 + x )}}{x^4 }.
    \label{eq:F_de_x}
\end{equation}
% ${\cal F}(x) = \frac{4 }{3 x} - \frac{2}{x^2} + \frac{4 }{x^3} - \frac{4\,\ln{(1 + x )}}{x^4 } $.
%
\\For $\textrm{Re}_0 \gg 1$, a Taylor expansion of equation~(\ref{eq:SPHERE_tau_CF_full}) gives $\tau_{\rm S}^{\rm conv}  \approx  {4\tau_{\rm S} }/({3 \beta_{\rm ev} \textrm{Re}_0^{1/2}\textrm{Sc}^{1/3}})$ meaning $\tau_{\rm S}^{\rm conv} \propto U_{\rm air}^{-1/2}$.

\subsubsection{Axisymmetric drop on a fiber}

Equation~(\ref{eq:SPHERE_tau_CF_full}) is only valid when the liquid temperature is constant with respect to $U_{\rm air}$ and equal to the temperature reached in diffusive regime. While spherical droplets temperature are minimally affected by air flow, the temperature gradient within the fiber, and consequently the heat flux $Q'_{\rm h}$ transferred by the solid to an axisymmetric drop on a fiber, depend on the air velocity. As air velocity increases, $Q'_{\rm h}$ diminishes, causing a decrease in liquid temperature, and it eventually becomes negligible in comparison to the heat exchange occurring between the droplet and the air at high Reynolds number~\cite{Shih1995, Shringi2013, Yang2001}.
Nonetheless, the variation of  $Q'_{\rm h}$ with $U_{\rm air}$ acts as a multiplicative factor which enables to write the lifetime of an axisymmetric drop on a fiber as:

\begin{equation}\label{eq:FIBER_tau_CF_full}
    \begin{cases}
        \tau_{\rm SF}^{\rm conv}  =  \tau_{\rm SF} &\text{ if } x = 0,\\
        \tau_{\rm SF}^{\rm conv}  \propto  \tau_{\rm SF}\, {\cal F}(x) &\text{ if } x > 0.
    \end{cases}
\end{equation}
where $\tau_{\rm SF}$ is obtained by numerically solving equation~(\ref{eq:eq_diff_R_de_t_sphere_on_fiber}).
The proportionality coefficient is expected to be close to unity.

\subsection{Results and Discussion}
\begin{figure}
    \centering
    \includegraphics[width = 1\linewidth]{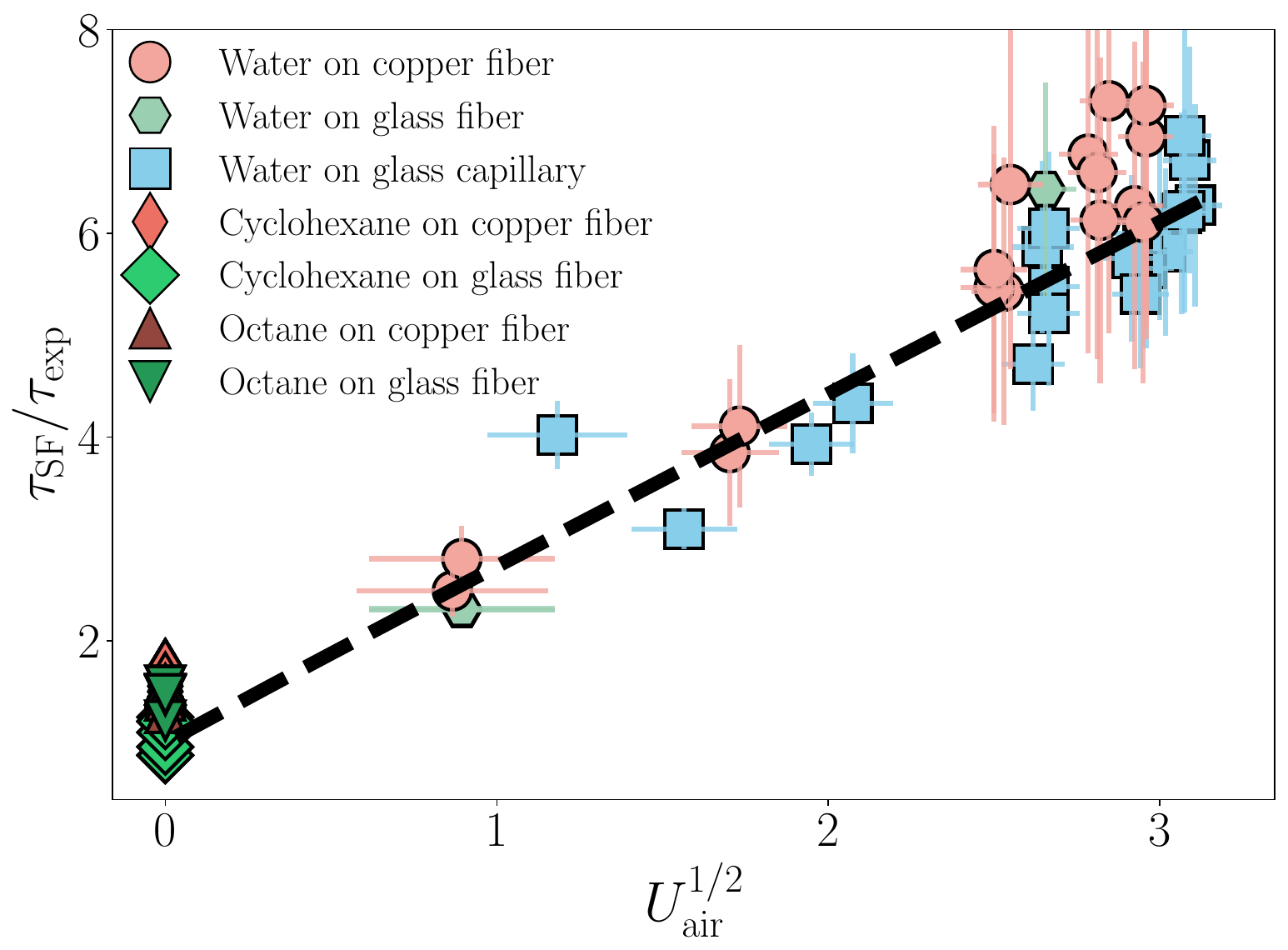}
    \caption{Lifetime ratio $\tau_{\rm SF}/\tau_{\rm exp}$ as a function of $U_{\rm air}^{1/2}$ where all the data presented in Figure~\ref{fig:timelapse_and_raw_data} are reported.
    The black dashed line represents equation~(\ref{eq:FIBER_tau_CF_full}) with a proportionality constant equal to 1.
    }
    \label{fig:5_collapse_tau_exp_vs_air_velo}
\end{figure}

All the experimental results were obtained for droplets initially in ``barrel'' morphology.
However, for $U_{\rm air} > 5 $~m/s, the droplet transitions into a ``blown'' droplet shape, as described by Bintein et al.~\cite{Bintein2015, Bintein2019}.
The study in~\cite{Bintein2015} shows that the flow induces a rotational motion of the droplet around the fiber, while the droplet itself undergoes minimal deformation due to the air flux.
These measurements notably indicate that the droplets maintain a consistent apparent surface area facing the wind, enabling a comparison of measurements conducted at both high and low velocities and the use of the spherical drop model.
Moreover in the experiments the drops are not self-propelled along the fiber even though is it expected for the considered Reynolds numbers~\cite{Bintein2019}.
This absence of self-propulsion might indicate that the flow is not perfectly laminar.

Figure~\ref{fig:5_collapse_tau_exp_vs_air_velo} shows the inverse of the experimental lifetimes normalized by the diffusive lifetime of a drop on a fiber $\tau_{\rm SF}$ as a function of $U_{\rm air}^{1/2}$.
As previously discussed, the experimental lifetimes are inversely proportional to the square root of the air velocity at high velocity.
Moreover, the experimental data points align, with a difference of the order of 10~\% between the model and the experimental results, with the dashed black line representing equation~(\ref{eq:FIBER_tau_CF_full}) (with a proportionality constant equals to 1).
This agreement is observed without the introduction of any adjustable parameters.

%%%%%%%%%%%%%%%%%%%%%
%
%%%%%%%%%%%%%%%%%%%%%
\section{Conclusion}\label{sec:conclusion}
A model is developed to predict the lifetime of a barrel-shaped drop on a fiber evaporating in both diffusive and convective regimes.
The system is modeled as a sphere pierced by a fiber which cools down while evaporating and exchanges heat with its environment.
In the diffusive regime, the heat fluxes exchanged between the liquid and the atmosphere on one hand, and the liquid and the fiber on the other hand were calculated to obtain the temperature of the drop, its evaporation rate, and its lifetime.

This analysis identifies a dimensionless number, $\Tilde{\mathcal{Q}}_{\rm fiber}$ that compares the heat flux transferred to the drop by the fiber to the heat flux exchanged between the liquid and the air. This number quantifies the role of the fiber in the heat exchange.
When $\Tilde{\mathcal{Q}}_{\rm fiber} \gg 1$, the liquid is thermalized by the fiber and the cooling of the drop is negligible, so the system can be accurately described by a model of an adiabatically evaporating sphere.
This is observed with metallic fibers, which are good thermal conductors.
For $\Tilde{\mathcal{Q}}_{\rm fiber} \ll 1$, the heat flux brought by the fiber to the drop is negligible. Therefore, the drop evaporates as an airborne sphere.
The amount of cooling depends solely on the physical properties of the liquid.
Here, we have shown that for macroscopic drops ($R > a$) placed on insulating fibers, we have $\Tilde{\mathcal{Q}}_{\rm fiber} \sim 1$.
This means that the heat flux brought by the fiber to the drop cannot be neglected and the drop cannot be considered as an isolated sphere in the air, even for good thermal insulating materials such as glass.
Additionally, the model offers a consistent representation that agrees reasonably well with the measured lifetimes in the diffusive regime for all experimentally tested liquid/solid couples.

Using the pioneering works of Fr\"ossling~\cite{Frossling1938} and Fuchs~{\cite{Fuchs1959}} associated with the model developed in the diffusive regime, we have shown that we are able to propose a semi-quantitative model of the lifetime of drops placed on fibers evaporating under an external air flow. In particular, the drop lifetime in forced-convective regime in proportional to its lifetime in diffusive regime and inversely proportional to the square-root of the air velocity.

We believe that this study, by understanding the role of the fiber on the heat transfer, can also provide new insights into the drying of droplets containing particles {\cite{Gelderblom2022, Larson2014, Hu2006,Brutin2022}}.
The present study opens up new possibilities for understanding particle deposits in more complex geometries during the evaporation of a drop.
For instance, as demonstrated in~{\cite{Corpart2023JFM}}, when a axisymmetric water droplet dries on a fiber, the resulting deposit is more uniform compared to when it dries on a flat surface.
Finally, one of the most important hypothesis in our models is the symmetry of the droplet on the fiber that is not verified experimentally due to gravity.
Future theoretical and numerical works will be useful to decipher the role of gravity on the evaporation dynamics {\cite{Gupta2021}}.

%%%%%%%%%%%%%%%%%%%%%
%
%%%%%%%%%%%%%%%%%%%%%
\section*{Conflicts of interest}
There are no conflicts of interest.

\section*{Acknowledgements}
We acknowledge Saint-Gobain and ANRT for funding this study.
We thank for the funding provided by CNRS Physique for the infrared camera.

\begin{center}
    \rule{0.5\linewidth}{0.4pt}
\end{center}

%%%%~~~~~~~~~~~~~~~~~~~~~~~~~%%%
%%%%~~~~~~~~~~~~~~~~~~~~~~~~~%%%
%%%%~~~~~~~APPENDIX~~~~~~~~~~%%%
%%%%~~~~~~~~~~~~~~~~~~~~~~~~~%%%
%%%%~~~~~~~~~~~~~~~~~~~~~~~~~%%%

%\newpage
% Format for appendix section titles
\titleformat{\section}[block]{\centering\bfseries}{Appendix \thesection:}{1em}{}
\titleformat{\subsection}{\centering\itshape}{\arabic{subsection}}{1em}{}

\appendix
\onecolumn
\counterwithin{figure}{section}
%%%%%%%%%%%%%%%%%%%%%
%
%%%%%%%%%%%%%%%%%%%%%
\section{Materials and methods}
% \TODO{TExt description fig Photos + schema soufflerie}
\subsection{Experimental setup}

\begin{figure}[h!]
    \centering
    \includegraphics[width = \linewidth]{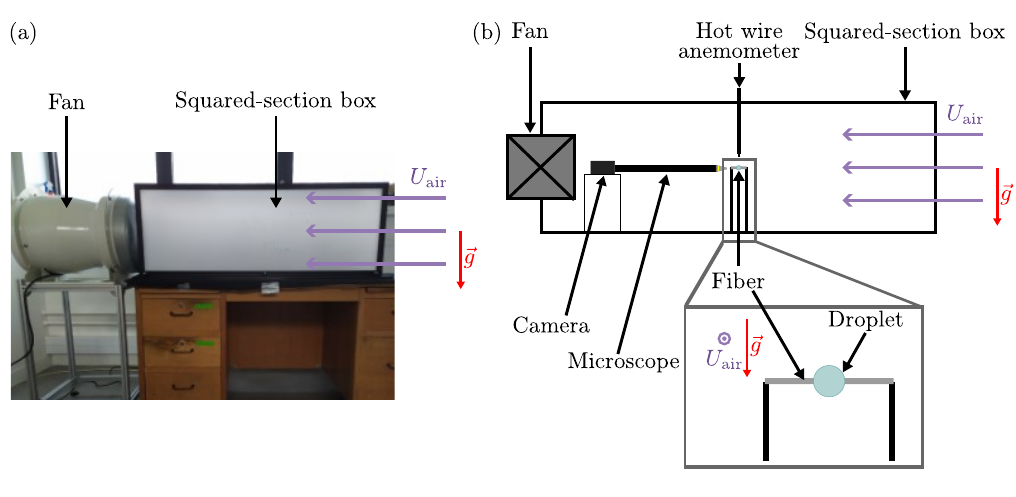}
    \caption{(a) Photograph and (b) schematic representation of the homemade wind tunnel.}
    \label{fig:SM_scheme_wind_tunnel}
\end{figure}

The experimental setup described in the main article, the homemade wind tunnel, is shown in more detail in Figure~\ref{fig:SM_scheme_wind_tunnel}. Figure~\ref{fig:SM_scheme_wind_tunnel}(a) shows a side view of the outside of the homemade wind tunnel and Figure~\ref{fig:SM_scheme_wind_tunnel}(b) provides a schematic view of the setup as well as a zoomed view of the drop placed on a fiber.

\subsection{Morphologies of the drops on the fibers}
\begin{figure}[h!]
    \centering
    \includegraphics[width = \linewidth]{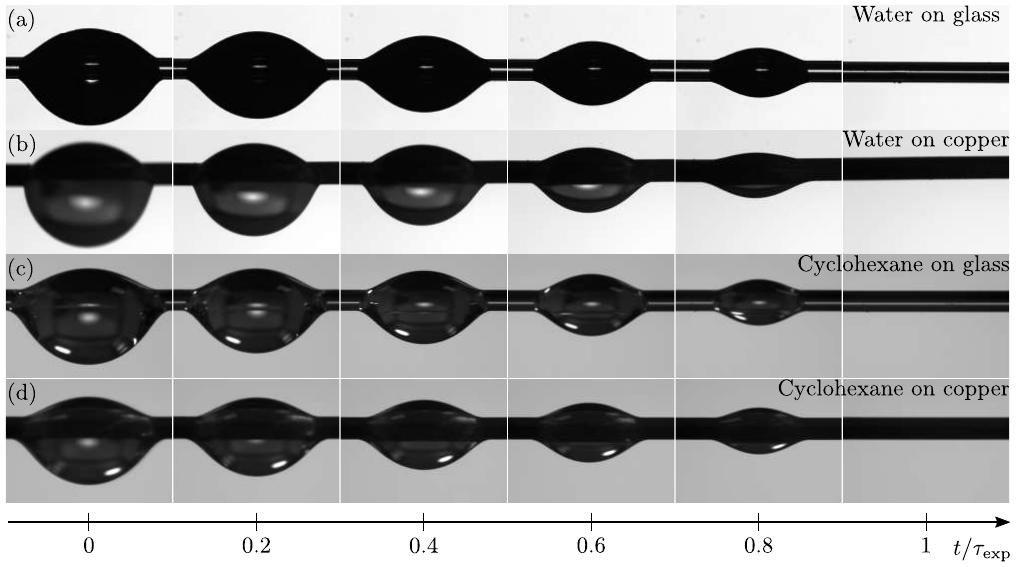}
    \caption{Time-lapse of the evaporation of a drop of initial volume $\Omega_0 \approx 0.8$~µL deposited on a fiber of radius $a = 125$~µm observed in side view. (a) Water on glass fiber (same photos as in Figure 1(a) of the main text.), (b) water on copper fiber, (c) cyclohexane on glass fiber and (d) cyclohexane on copper fiber}
    \label{fig:SM_timelapse_for_all_liq}
\end{figure}

All the lifetime measurements were obtained for drops adopting initially an axisymetric configuration, known as the barrell shape~\cite{Carroll1976, Chou2011, Mchale2002}, on the fiber. The timelapses of a water drop on a glass (Fig.~\ref{fig:SM_timelapse_for_all_liq}(a)) and a copper fiber (Fig.~\ref{fig:SM_timelapse_for_all_liq}(b)) and a cyclohexane drop on a glass (Fig.~\ref{fig:SM_timelapse_for_all_liq}(c)) and a copper fiber (Fig.~\ref{fig:SM_timelapse_for_all_liq}(d)) evaporating in diffusive regime ($U_{\rm air} = 0$) are shown in Figure~\ref{fig:SM_timelapse_for_all_liq}. Figure~\ref{fig:SM_timelapse_for_all_liq} shows that all the liquids adopt a barrel configuratin on the tested fiber and the different wetting conditions have a small impact on the geometry of the liquid as expected for drop place on a fiber~\cite{Carroll1976, Chou2011, Mchale2002}.
% \TODO{Text Photos des differents liquides sur les fibres pour montrer morpho}

All lifetime measurements were obtained for drops that initially adopt an axisymmetric configuration on the fiber, known as the barrel shape~\cite{Carroll1976, Chou2011, Mchale2002}. The time lapses of a water drop on glass  (Fig.~\ref{fig:SM_timelapse_for_all_liq}(a)) and copper fiber (Fig.~\ref{fig:SM_timelapse_for_all_liq}(b)) and a cyclohexane drop on glass (Fig.~\ref{fig:SM_timelapse_for_all_liq}(c)) and copper fiber (Fig.~\ref{fig:SM_timelapse_for_all_liq}(d)) evaporating in the diffusive regime ($U_{\rm air} = 0$) are shown in Figure~\ref{fig:SM_timelapse_for_all_liq}. Figure~\ref{fig:SM_timelapse_for_all_liq} shows that all liquids adopt a barrel configuration, kept throughout the evaporation, on the tested fiber and that the different wetting conditions have little effect on the geometry of the liquid, as expected for drops placed on a fiber~\cite{Carroll1976, Chou2011, Mchale2002}.

%%%%%%%%%%%%%%%%%%%%%
%
%%%%%%%%%%%%%%%%%%%%%
\newpage
\section{Physicochemical properties of the studied liquids}\label{sec:APPENDIX-properties}

\begin{figure}[h!]
    \centering
    \includegraphics[width = .8\linewidth]{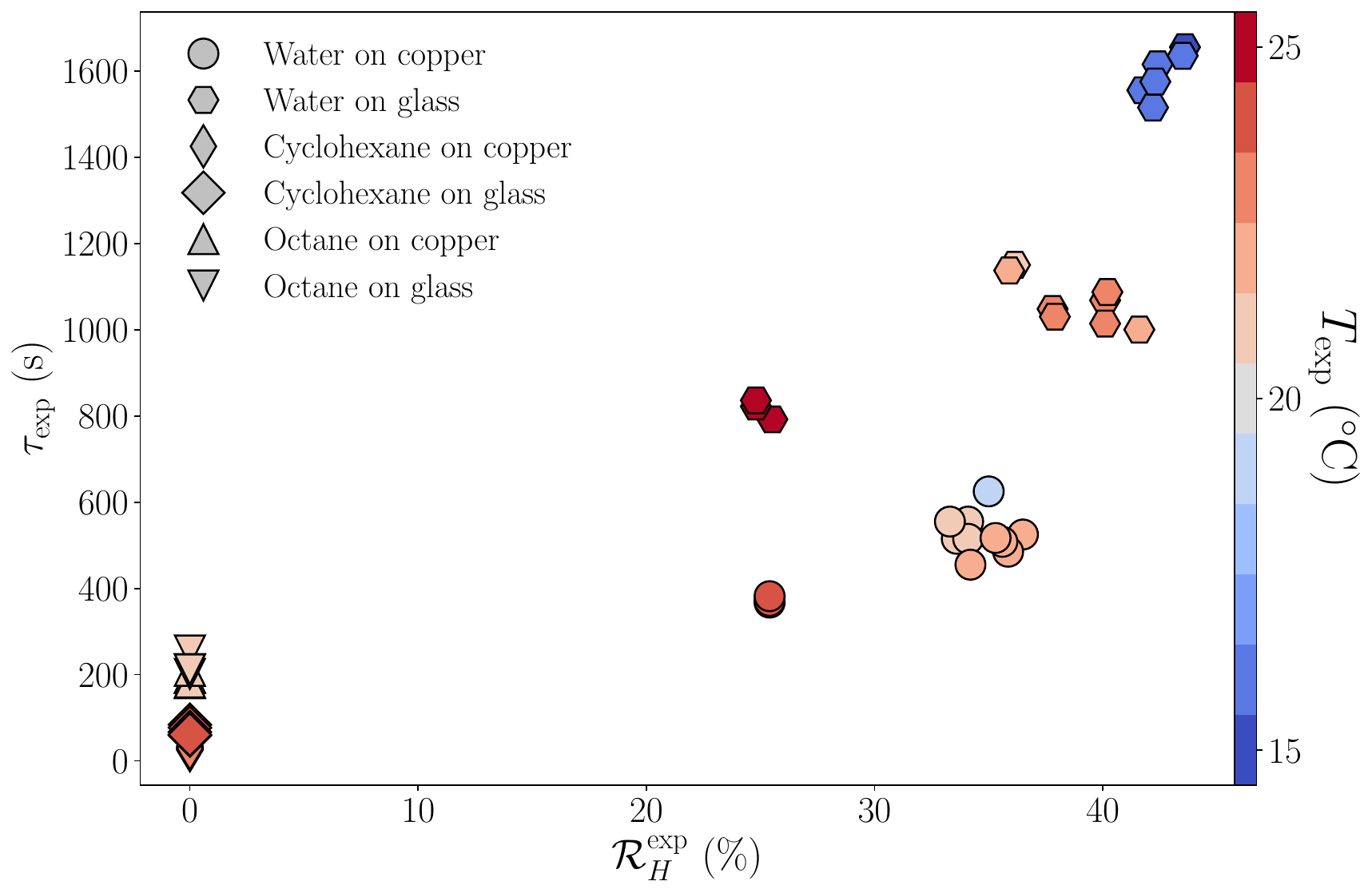}
    \caption{Evolution of the measured lifetimes in diffusive regime ($U_{\rm air} = 0$) as a function of the ambient humidity and temperature conditions.}
    \label{fig:SM_tau_exp_vs_RH_exp_all_DS_data}
\end{figure}

In our experiments, temperature and relative humidity are variable.
In Figure~\ref{fig:SM_tau_exp_vs_RH_exp_all_DS_data}, the lifetimes measured in the diffusive regime are plotted as a function of the relative humidity $\mathcal{R}_{\rm H}^{\rm exp}$. For water droplets, relative humidity is measured using a commercial hygrometer. Since there is no cyclohexane or octane vapor in the atmosphere, $\mathcal{R}_{\rm H}^{\rm exp} = 0$ for the alkanes. The color of the markers provides information about the air temperature away from the drop, $T_{\rm exp}$, during the measurement of the lifetime of each individual drop.

% \TODO{Raw experimental data for $U_{\rm air} = 0$}

The air temperature varies between $15~\rm ^\circ C$ and $25~\rm ^\circ C$, which has a significant effect on the evaporation rate of the liquid and precludes direct comparison of results obtained for a given liquid/solid system at a constant relative humidity. This can be observed, for example, in~\ref{fig:SM_tau_exp_vs_RH_exp_all_DS_data}, where the data points representing the results obtained for a drop of water on a copper fiber (circles) at relative humidity $\mathcal{R}_{\rm H} \approx 35~\rm \%$ are not superimposed because the temperature varies between $19$ and $22~\rm ^\circ C$.

We thus need to calculate the theoretical lifetimes, predicted by our model, in the experimental conditions. 
To do so we need to prescribe the variation of several physical constant with the temperature as described in details in~\cite{Corpart2023}. Here we summarize the phenomenological equations used to calculate the relevant parameters as well as some data extracted from the literature.
We will consider the example of an isolated spherical droplet whose lifetime is determined by the combination of equations~(2) and (6) of the main article.
To calculate the lifetime for a given $T_\infty$, we need the values of $\rho$, $\chi(T_\infty) = \Delta_{\rm vap} H {\cal D}(T_\infty) c_{\rm sat}(T_\infty)/  \lambda_{\rm air}(T_\infty)$, $\alpha_1(T_\infty)$ and $\alpha_2(T_\infty)$. We limit this study to ambient temperatures range, $T_\infty \in [0, 30]~^\circ \rm C$.  In the following we review the temperature variations of all the submentioned parameters.
\subsection{Density and enthalpy of vaporization}
In the selected temperature range, the values of  $\rho$ and $\Delta_{\rm vap} H$  remain almost constant. 
Therefore, we have chosen to use their values at 20 and 25~$^\circ \rm C$, respectively, as these are the values commonly found in handbooks of physics. 
These data are gathered in the Table~\ref{tab:APPENDIX-C_constantes_physiques_liq} for all the liquids studied.

\begin{table}[h!]
\centering
\begin{tabular}{|c|c|c|c|c|}
\hline
%\rowcolor[HTML]{65A8A5} 
                               & Water   & Cyclohexane & Octane & Ref.       \\ \hline
$M$ (kg/mol)   & $18 \times 10^{-3}$  & $84 \times 10^{-3}$  & $114 \times 10^{-3}$ & \cite{Lide2008} \\ \hline
$\Delta_{\rm vap} H$ (J/kg) & $2.44 \times 10^{6}$ & $3.93 \times 10^{5}$ & $3.64 \times 10^{5}$ & \cite{Lide2008} \\ \hline
$\rho$ (kg/m$^3$)              & 998.2 & 779         & 702.5  & \cite{Lide2008} \\ \hline
$V$~(cm$^3$/mol)          & 13.1  & 104.8       & 168.7  & \cite{Fuller1969} \\ \hline
\end{tabular}
\caption{Physical constants of the liquids under consideration, $M$ is the molar mass and $V$ the molecular volume as described in~\cite{Fuller1969}. 
The mass density $\rho$ is given for the liquid at $20~^\circ$C, while the enthalpy of vaporization $\Delta_{\rm vap} H$ is specified at 25~$^\circ$C.}
\label{tab:APPENDIX-C_constantes_physiques_liq}
\end{table}

\subsection{Diffusion coefficient $\cal D$}
The value of the diffusion coefficient of vapor in air at $T_\infty$ is estimated by the Fuller, Schettler, and Giddings' method~\cite{Fuller1966, Fuller1969} which writes for a molecule $A$ diffusing in $B$
\begin{equation}\label{eq:APPENDIX-Fuller_equation}
    \mathcal{D}(A,B) = \frac{ T^{1.75}\sqrt{\frac{1}{M_{\rm A}} + \frac{1}{M_{\rm B}}}}{P_{\rm atm}  \left ( V_A^{1/3} + V_B^{1/3}\right )^2} \cdot 10^{-6}.
\end{equation}
The diffusion coefficient ${\cal D}$ is expressed in m$^2$/s, the molar mass of compound $i$ is in g/mol and $V_i$ is the diffusion volume of the molecule $i$ where $V_i = \sum_{j} n_j V_j$ with $j$ a given atom composing the molecule~\cite{Fuller1969}. The diffusion volume and molar masses of the molecules are given in Table~\ref{tab:APPENDIX-C_constantes_physiques_liq} for the considered liquids.

\subsection{Saturating vapor concentration $c_{\rm sat}$ at $T_\infty$}

To obtain the value of $c_{\rm sat}(T_\infty)$, vapor is treated as an ideal gas, so the saturated vapor concentration $c_{\rm sat}(T)$ is given by:

\begin{equation}\label{eq:APPENDIX-C_c_sat_ideal_gas}
c_{\rm sat}(T) = \frac{p_{\rm sat}(T) M}{ {\cal R} T},
\end{equation}
where $p_{\rm sat}$ represents the saturated vapor pressure, $M$ is the molar mass of the vapor, ${\cal R}$ is the ideal gas constant, and $T$ is the temperature in Kelvin.
The temperature dependence of $p_{\rm sat}$ is approximated by Antoine's equation.
\begin{equation}\label{eq:APPENDIX-C_Antoine_equation}
    p_{\rm sat}(T) = P^\circ\, 10^{A - \frac{B}{C + T}},
\end{equation}
where $ P^\circ = 10^3$~Pa and $T$ is in Kelvin.

The constants, $A$, $B$, and $C$, are determined by fitting the data extracted from literature for the considered liquids which are presented in Table~\ref{tab:APPENDIX-C_data_psat_vs_T}, with the Antoine's equation (Eq.~(\ref{eq:APPENDIX-C_Antoine_equation})). 
Table~\ref{tab:APPENDIX-C_coeff_Antoine} gathers the coefficients obtained for water, cyclohexane, and octane within the temperature range of $T \in [0, 30]~^\circ$C.

\begin{table}[h!]
\centering
\begin{tabular}{c|c|c|c|}
\cline{2-4}
                                  & $A$    & $B$    & $C$     \\ \hline
\multicolumn{1}{|c|}{Water}       & 7.34 & 1808 & -33.9 \\ \hline
\multicolumn{1}{|c|}{Cyclohexane} & 6.17 & 1304 & -40.4 \\ \hline
\multicolumn{1}{|c|}{Octane}      & 11.2 & 4791 & 139   \\ \hline
\end{tabular}
\caption{Antoine's coefficients. The coefficients $A$, $B$, and $C$ have been obtained by fitting the data extracted from the literature with the Antoine equation (Eq.~(\ref{eq:APPENDIX-C_Antoine_equation})). 
Coefficients $B$ and $C$ are in Kelvin.
The data are available in Table \ref{tab:APPENDIX-C_data_psat_vs_T}. 
The coefficients given here are valid for temperatures in the range of $T \in [0, 30]^\circ$C.}
\label{tab:APPENDIX-C_coeff_Antoine}
\end{table}

\begin{table}
\centering

\label{tab:APPENDIX-C_data_psat_vs_T}
\resizebox{\linewidth}{!}{%
\begin{tabular}{|c|c|c|c|c|c|c|c|c|c|c|c|c|c|c|} \hline
\multicolumn{3}{|c|}{Water} & \multicolumn{6}{c|}{Cyclohexane} & \multicolumn{6}{c|}{Octane} \\ \hline
$T$ (°C) & $p_{\rm sat}$ (Pa) & Ref & $T$ (°C) & $p_{\rm sat}$ (Pa) & Ref & $T$ (°C) & $p_{\rm sat}$ (Pa) & Ref & $T$ (°C) & $p_{\rm sat}$ (Pa) & Ref & $T$ (°C) & $p_{\rm sat}$ (Pa) & Ref \\ \hline
0 & $6.11 \times 10^{2}$ & \cite{Lide2008} & 5.25 & $4.93 \times 10^{3}$ & \cite{Jakli1978} & 20.21 & $1.04 \times 10^{4}$ & \cite{Jakli1978} & 0 & $3.60 \times 10^{2}$ & \cite{Cook1958} & 25.7 & $2.00 \times 10^{3}$ & \cite{Young1928} \\ \hline
1 & $6.57 \times 10^{2}$ & \cite{Lide2008} & 5.26 & $4.95 \times 10^{3}$ & \cite{Jakli1978} & 20.36 & $1.05 \times 10^{4}$ & \cite{Jakli1978} & 0 & $3.87 \times 10^{2}$ & \cite{Cook1958} & 26.35 & $2.00 \times 10^{3}$ & \cite{Young1928} \\ \hline
2 & $7.06 \times 10^{2}$ & \cite{Lide2008} & 6.56 & $5.33 \times 10^{3}$ & \cite{Jakli1978} & 21.53 & $1.11 \times 10^{4}$ & \cite{Jakli1978} & 0 & $4.33 \times 10^{2}$ & \cite{Linder1931} & 26.75 & $2.04 \times 10^{3}$ & \cite{Young1900} \\ \hline
3 & $7.58 \times 10^{2}$ & \cite{Lide2008} & 7.27 & $5.49 \times 10^{3}$ & \cite{Jakli1978} & 21.64 & $1.11 \times 10^{4}$ & \cite{Jakli1978} & 0 & $5.33 \times 10^{2}$ & \cite{Linder1931} & 27.4 & $2.13 \times 10^{3}$ & \cite{Dejoz1996} \\ \hline
4 & $8.14 \times 10^{2}$ & \cite{Lide2008} & 7.46 & $5.54 \times 10^{3}$ & \cite{Jakli1978} & 22.608 & $1.17 \times 10^{4}$ & \cite{Ewing2000} & 0 & $3.93 \times 10^{2}$ & \cite{Linder1931} & 29.6 & $2.41 \times 10^{3}$ & \cite{Dejoz1996} \\ \hline
5 & $8.73 \times 10^{2}$ & \cite{Lide2008} & 7.62 & $5.47 \times 10^{3}$ & \cite{Jakli1978} & 23.363 & $1.21 \times 10^{4}$ & \cite{Ewing2000} & 1.2 & $4.20 \times 10^{2}$ & \cite{Young1900} & 29.65 & $2.41 \times 10^{3}$ & \cite{Young1900} \\ \hline
6 & $9.35 \times 10^{2}$ & \cite{Lide2008} & 8.188 & $5.78 \times 10^{3}$ & \cite{Ewing2000} & 24.5 & $1.27 \times 10^{4}$ & \cite{Jakli1978} & 3.7 & $4.87 \times 10^{2}$ & \cite{Linder1931} & 30 & $2.47 \times 10^{3}$ & \cite{Weiguo1990} \\ \hline
7 & $1.00 \times 10^{3}$ & \cite{Lide2008} & 8.23 & $5.78 \times 10^{3}$ & \cite{Jakli1978} & 24.99 & $1.30 \times 10^{4}$ & \cite{Jakli1978} & 4.4 & $5.27 \times 10^{2}$ & \cite{Young1900} &  &  &  \\ \hline
8 & $1.07 \times 10^{3}$ & \cite{Lide2008} & 8.59 & $5.88 \times 10^{3}$ & \cite{Jakli1978} & 25 & $1.30 \times 10^{4}$ & \cite{Lide2008} & 5.26 & $5.32 \times 10^{2}$ & \cite{Carruth1973} &  &  &  \\ \hline
9 & $1.15 \times 10^{3}$ & \cite{Lide2008} & 9.08 & $6.06 \times 10^{3}$ & \cite{Jakli1978} & 25 & $1.30 \times 10^{4}$ & \cite{Bell1968} & 8.2 & $6.53 \times 10^{2}$ & \cite{Young1900} &  &  &  \\ \hline
10 & $1.23 \times 10^{3}$ & \cite{Lide2008} & 9.698 & $6.24 \times 10^{3}$ & \cite{Ewing2000} & 25 & $1.30 \times 10^{4}$ & \cite{Cruickshank1967} & 9.25 & $7.13 \times 10^{2}$ & \cite{Young1900} &  &  &  \\ \hline
11 & $1.31 \times 10^{3}$ & \cite{Lide2008} & 11.06 & $6.68 \times 10^{3}$ & \cite{Jakli1978} & 25 & $1.30 \times 10^{4}$ & \cite{Ksiazczak1991} & 9.55 & $7.33 \times 10^{2}$ & \cite{Young1900} &  &  &  \\ \hline
12 & $1.40 \times 10^{3}$ & \cite{Lide2008} & 11.13 & $6.70 \times 10^{3}$ & \cite{Jakli1978} & 25 & $1.30 \times 10^{4}$ & \cite{Washburn1935} & 11.2 & $8.20 \times 10^{2}$ & \cite{Young1900} &  &  &  \\ \hline
13 & $1.50 \times 10^{3}$ & \cite{Lide2008} & 11.25 & $6.74 \times 10^{3}$ & \cite{Jakli1978} & 26.48 & $1.39 \times 10^{4}$ & \cite{Jakli1978} & 12.55 & $8.87 \times 10^{2}$ & \cite{Young1900} &  &  &  \\ \hline
14 & $1.60 \times 10^{3}$ & \cite{Lide2008} & 11.356 & $6.79 \times 10^{3}$ & \cite{Ewing2000} & 27.42 & $1.45 \times 10^{4}$ & \cite{Jakli1978} & 14.35 & $9.93 \times 10^{2}$ & \cite{Young1900} &  &  &  \\ \hline
15 & $1.71 \times 10^{3}$ & \cite{Lide2008} & 11.52 & $6.84 \times 10^{3}$ & \cite{Jakli1978} & 28.08 & $1.49 \times 10^{4}$ & \cite{Jakli1978} & 14.4 & $1.00 \times 10^{3}$ & \cite{Lide2008} &  &  &  \\ \hline
16 & $1.82 \times 10^{3}$ & \cite{Lide2008} & 12.17 & $7.07 \times 10^{3}$ & \cite{Jakli1978} & 28.346 & $1.51 \times 10^{4}$ & \cite{Ewing2000} & 16.41 & $1.05 \times 10^{3}$ & \cite{Carruth1973} &  &  &  \\ \hline
17 & $1.94 \times 10^{3}$ & \cite{Lide2008} & 12.88 & $7.32 \times 10^{3}$ & \cite{Jakli1978} & 28.49 & $1.52 \times 10^{4}$ & \cite{Jakli1978} & 17.05 & $1.15 \times 10^{3}$ & \cite{Young1900} &  &  &  \\ \hline
18 & $2.06 \times 10^{3}$ & \cite{Lide2008} & 14 & $7.73 \times 10^{3}$ & \cite{Jakli1978} & 28.97 & $1.55 \times 10^{4}$ & \cite{Jakli1978} & 18.1 & $1.25 \times 10^{3}$ & \cite{Dejoz1996} &  &  &  \\ \hline
19 & $2.20 \times 10^{3}$ & \cite{Lide2008} & 14.39 & $7.88 \times 10^{3}$ & \cite{Jakli1978} & 30 & $1.62 \times 10^{4}$ & \cite{Carmona2000} & 19.7 & $1.37 \times 10^{3}$ & \cite{Young1900} &  &  &  \\ \hline
20 & $2.34 \times 10^{3}$ & \cite{Lide2008} & 14.59 & $7.98 \times 10^{3}$ & \cite{Jakli1978} &  &  &  & 20.05 & $1.47 \times 10^{3}$ & \cite{Young1928} &  &  &  \\ \hline
21 & $2.49 \times 10^{3}$ & \cite{Lide2008} & 14.97 & $8.12 \times 10^{3}$ & \cite{Jakli1978} &  &  &  & 20.4 & $1.47 \times 10^{3}$ & \cite{Young1928} &  &  &  \\ \hline
22 & $2.64 \times 10^{3}$ & \cite{Lide2008} & 15.07 & $8.15 \times 10^{3}$ & \cite{Jakli1978} &  &  &  & 20.7 & $1.45 \times 10^{3}$ & \cite{Dejoz1996} &  &  &  \\ \hline
23 & $2.81 \times 10^{3}$ & \cite{Lide2008} & 15.215 & $8.23 \times 10^{3}$ & \cite{Ewing2000} &  &  &  & 20.9 & $1.47 \times 10^{3}$ & \cite{Young1928} &  &  &  \\ \hline
24 & $2.99 \times 10^{3}$ & \cite{Lide2008} & 17.163 & $9.04 \times 10^{3}$ & \cite{Ewing2000} &  &  &  & 23.15 & $1.65 \times 10^{3}$ & \cite{Young1900} &  &  &  \\ \hline
25 & $3.17 \times 10^{3}$ & \cite{Lide2008} & 17.21 & $9.04 \times 10^{3}$ & \cite{Jakli1978} &  &  &  & 23.96 & $1.57 \times 10^{3}$ & \cite{Carruth1973} &  &  &  \\ \hline
26 & $3.36 \times 10^{3}$ & \cite{Lide2008} & 18.18 & $9.48 \times 10^{3}$ & \cite{Jakli1978} &  &  &  & 24.6 & $1.82 \times 10^{3}$ & \cite{Dejoz1996} &  &  &  \\ \hline
27 & $3.57 \times 10^{3}$ & \cite{Lide2008} & 18.75 & $9.74 \times 10^{3}$ & \cite{Jakli1978} &  &  &  & 25 & $1.86 \times 10^{3}$ & \cite{Lide2008} &  &  &  \\ \hline
28 & $3.78 \times 10^{3}$ & \cite{Lide2008} & 19.3 & $1.00 \times 10^{4}$ & \cite{Lide2008} &  &  &  & 25 & $1.86 \times 10^{3}$ & \cite{Bell1968} &  &  &  \\ \hline
29 & $4.01 \times 10^{3}$ & \cite{Lide2008} & 19.878 & $1.03 \times 10^{4}$ & \cite{Ewing2000} &  &  &  & 25 & $1.85 \times 10^{3}$ & \cite{Cook1958} &  &  &  \\ \hline
30 & $4.25 \times 10^{3}$ & \cite{Lide2008} & 20.11 & $1.04 \times 10^{4}$ & \cite{Jakli1978} &  &  &  & 25 & $1.87 \times 10^{3}$ & \cite{Weiguo1990} &  &  &  \\ \hline
\end{tabular}
}
\caption{Saturated vapor pressures $p_{\rm sat}$ as a function of temperature $T$ for water, cyclohexane, and octane.}
\end{table}

\subsection{Values of $\alpha_1$ and $\alpha_2$}
To perform analytically the calculation of droplet evaporation we choose to use an approximation to describe $c_{\rm sat}(T)$. To model the temperature-dependent behavior of the saturation concentration, we employ a quadratic approximation (Eq.~(5) of the main text), a method previously used in Ref.~\cite{Corpart2023}.
The coefficients $\alpha_1$ and $\alpha_2$ are determined for a given air temperature $T_\infty$, by fitting the $c_{\rm sat}$ values obtained from the literature (Table~\ref{tab:APPENDIX-C_data_psat_vs_T} combined with Eq.~(\ref{eq:APPENDIX-C_c_sat_ideal_gas})).
The resulting values are compiled in Table~\ref{tab:APPENDIX-C_alpha_1_et_alpha_2} for temperatures between 10 and 30~$^\circ \rm C$.

\begin{table}[]
\centering
\resizebox{\textwidth}{!}{%
\begin{tabular}{c|cc|cc|cc|}
\cline{2-7}
 &
  \multicolumn{2}{c|}{Water} &
  \multicolumn{2}{c|}{Cyclohexane} &
  \multicolumn{2}{c|}{Octane} \\ \hline
\multicolumn{1}{|c|}{$T_\infty$ ($^\circ$C)} &
  \multicolumn{1}{c|}{$\alpha_1$} &
  $\alpha_2$ &
  \multicolumn{1}{c|}{$\alpha_1$} &
  $\alpha_2$ &
  \multicolumn{1}{c|}{$\alpha_1$} &
  $\alpha_2$ \\ \hline
\multicolumn{1}{|c|}{10} &
  \multicolumn{1}{c|}{$-6.24\times 10^{-2}$} &
  $1.41\times 10^{-3}$ &
  \multicolumn{1}{c|}{$-4.91\times 10^{-2}$} &
  $1.20\times 10^{-3}$ &
  \multicolumn{1}{c|}{$-7.40\times 10^{-2}$} &
  $3.14\times 10^{-3}$ \\ \hline
\multicolumn{1}{|c|}{11} &
  \multicolumn{1}{c|}{$-6.17\times 10^{-2}$} &
  $1.36\times 10^{-3}$ &
  \multicolumn{1}{c|}{$-4.78\times 10^{-2}$} &
  $9.64\times 10^{-4}$ &
  \multicolumn{1}{c|}{$-7.04\times 10^{-2}$} &
  $2.60\times 10^{-3}$ \\ \hline
\multicolumn{1}{|c|}{12} &
  \multicolumn{1}{c|}{$-6.11\times 10^{-2}$} &
  $1.32\times 10^{-3}$ &
  \multicolumn{1}{c|}{$-4.70\times 10^{-2}$} &
  $8.55\times 10^{-4}$ &
  \multicolumn{1}{c|}{$-6.72\times 10^{-2}$} &
  $2.19\times 10^{-3}$ \\ \hline
\multicolumn{1}{|c|}{13} &
  \multicolumn{1}{c|}{$-6.04\times 10^{-2}$} &
  $1.27\times 10^{-3}$ &
  \multicolumn{1}{c|}{$-4.62\times 10^{-2}$} &
  $7.88\times 10^{-4}$ &
  \multicolumn{1}{c|}{$-6.45\times 10^{-2}$} &
  $1.89\times 10^{-3}$ \\ \hline
\multicolumn{1}{|c|}{14} &
  \multicolumn{1}{c|}{$-5.97\times 10^{-2}$} &
  $1.23\times 10^{-3}$ &
  \multicolumn{1}{c|}{$-4.56\times 10^{-2}$} &
  $7.48\times 10^{-4}$ &
  \multicolumn{1}{c|}{$-6.24\times 10^{-2}$} &
  $1.66\times 10^{-3}$ \\ \hline
\multicolumn{1}{|c|}{15} &
  \multicolumn{1}{c|}{$-5.91\times 10^{-2}$} &
  $1.19\times 10^{-3}$ &
  \multicolumn{1}{c|}{$-4.51\times 10^{-2}$} &
  $7.20\times 10^{-4}$ &
  \multicolumn{1}{c|}{$-6.05\times 10^{-2}$} &
  $1.49\times 10^{-3}$ \\ \hline
\multicolumn{1}{|c|}{16} &
  \multicolumn{1}{c|}{$-5.84\times 10^{-2}$} &
  $1.15\times 10^{-3}$ &
  \multicolumn{1}{c|}{$-4.46\times 10^{-2}$} &
  $6.95\times 10^{-4}$ &
  \multicolumn{1}{c|}{$-5.89\times 10^{-2}$} &
  $1.35\times 10^{-3}$ \\ \hline
\multicolumn{1}{|c|}{17} &
  \multicolumn{1}{c|}{$-5.77\times 10^{-2}$} &
  $1.11\times 10^{-3}$ &
  \multicolumn{1}{c|}{$-4.42\times 10^{-2}$} &
  $6.76\times 10^{-4}$ &
  \multicolumn{1}{c|}{$-5.79\times 10^{-2}$} &
  $1.26\times 10^{-3}$ \\ \hline
\multicolumn{1}{|c|}{18} &
  \multicolumn{1}{c|}{$-5.71\times 10^{-2}$} &
  $1.07\times 10^{-3}$ &
  \multicolumn{1}{c|}{$-4.37\times 10^{-2}$} &
  $6.57\times 10^{-4}$ &
  \multicolumn{1}{c|}{$-5.73\times 10^{-2}$} &
  $1.20\times 10^{-3}$ \\ \hline
\multicolumn{1}{|c|}{19} &
  \multicolumn{1}{c|}{$-5.64\times 10^{-2}$} &
  $1.04\times 10^{-3}$ &
  \multicolumn{1}{c|}{$-4.33\times 10^{-2}$} &
  $6.40\times 10^{-4}$ &
  \multicolumn{1}{c|}{$-5.64\times 10^{-2}$} &
  $1.14\times 10^{-3}$ \\ \hline
\multicolumn{1}{|c|}{20} &
  \multicolumn{1}{c|}{$-5.58\times 10^{-2}$} &
  $1.01\times 10^{-3}$ &
  \multicolumn{1}{c|}{$-4.28\times 10^{-2}$} &
  $6.23\times 10^{-4}$ &
  \multicolumn{1}{c|}{$-5.56\times 10^{-2}$} &
  $1.08\times 10^{-3}$ \\ \hline
\multicolumn{1}{|c|}{21} &
  \multicolumn{1}{c|}{$-5.52\times 10^{-2}$} &
  $9.75\times 10^{-4}$ &
  \multicolumn{1}{c|}{$-4.24\times 10^{-2}$} &
  $6.06\times 10^{-4}$ &
  \multicolumn{1}{c|}{$-5.44\times 10^{-2}$} &
  $1.01\times 10^{-3}$ \\ \hline
\multicolumn{1}{|c|}{22} &
  \multicolumn{1}{c|}{$-5.45\times 10^{-2}$} &
  $9.45\times 10^{-4}$ &
  \multicolumn{1}{c|}{$-4.20\times 10^{-2}$} &
  $5.92\times 10^{-4}$ &
  \multicolumn{1}{c|}{$-5.34\times 10^{-2}$} &
  $9.56\times 10^{-4}$ \\ \hline
\multicolumn{1}{|c|}{23} &
  \multicolumn{1}{c|}{$-5.39\times 10^{-2}$} &
  $9.16\times 10^{-4}$ &
  \multicolumn{1}{c|}{$-4.16\times 10^{-2}$} &
  $5.79\times 10^{-4}$ &
  \multicolumn{1}{c|}{$-5.26\times 10^{-2}$} &
  $9.10\times 10^{-4}$ \\ \hline
\multicolumn{1}{|c|}{24} &
  \multicolumn{1}{c|}{$-5.33\times 10^{-2}$} &
  $8.88\times 10^{-4}$ &
  \multicolumn{1}{c|}{$-4.12\times 10^{-2}$} &
  $5.66\times 10^{-4}$ &
  \multicolumn{1}{c|}{$-5.19\times 10^{-2}$} &
  $8.74\times 10^{-4}$ \\ \hline
\multicolumn{1}{|c|}{25} &
  \multicolumn{1}{c|}{$-5.27\times 10^{-2}$} &
  $8.62\times 10^{-4}$ &
  \multicolumn{1}{c|}{$-4.09\times 10^{-2}$} &
  $5.52\times 10^{-4}$ &
  \multicolumn{1}{c|}{$-5.15\times 10^{-2}$} &
  $8.52\times 10^{-4}$ \\ \hline
\multicolumn{1}{|c|}{26} &
  \multicolumn{1}{c|}{$-5.21\times 10^{-2}$} &
  $8.36\times 10^{-4}$ &
  \multicolumn{1}{c|}{$-4.05\times 10^{-2}$} &
  $5.39\times 10^{-4}$ &
  \multicolumn{1}{c|}{$-5.10\times 10^{-2}$} &
  $8.28\times 10^{-4}$ \\ \hline
\multicolumn{1}{|c|}{27} &
  \multicolumn{1}{c|}{$-5.15\times 10^{-2}$} &
  $8.12\times 10^{-4}$ &
  \multicolumn{1}{c|}{$-4.01\times 10^{-2}$} &
  $5.27\times 10^{-4}$ &
  \multicolumn{1}{c|}{$-5.05\times 10^{-2}$} &
  $8.04\times 10^{-4}$ \\ \hline
\multicolumn{1}{|c|}{28} &
  \multicolumn{1}{c|}{$-5.09\times 10^{-2}$} &
  $7.88\times 10^{-4}$ &
  \multicolumn{1}{c|}{$-3.98\times 10^{-2}$} &
  $5.16\times 10^{-4}$ &
  \multicolumn{1}{c|}{$-5.01\times 10^{-2}$} &
  $7.82\times 10^{-4}$ \\ \hline
\multicolumn{1}{|c|}{29} &
  \multicolumn{1}{c|}{$-5.04\times 10^{-2}$} &
  $7.66\times 10^{-4}$ &
  \multicolumn{1}{c|}{$-3.94\times 10^{-2}$} &
  $5.05\times 10^{-4}$ &
  \multicolumn{1}{c|}{$-4.96\times 10^{-2}$} &
  $7.62\times 10^{-4}$ \\ \hline
\multicolumn{1}{|c|}{30} &
  \multicolumn{1}{c|}{$-4.98\times 10^{-2}$} &
  $7.44\times 10^{-4}$ &
  \multicolumn{1}{c|}{$-3.91\times 10^{-2}$} &
  $4.94\times 10^{-4}$ &
  \multicolumn{1}{c|}{$-4.92\times 10^{-2}$} &
  $7.42\times 10^{-4}$ \\ \hline
\end{tabular}%
}
\caption{Values of $\alpha_1$ and $\alpha_2$ for various temperatures $T_\infty$ for water, cyclohexane, and octane.}
\label{tab:APPENDIX-C_alpha_1_et_alpha_2}
\end{table}

\subsection{Temperature variation of air thermal conductivity $\lambda_{\rm air}$}

To calculate the value of $\lambda_{\rm air}$ at $T_\infty$ we use Andreas~\cite{Andreas1995} phenomenological equation for dry air, 
\begin{equation}\label{eq:APPENDIX-lambda_air_vs_T_Andreas95}
    \lambda_{\rm air}= -3.47 \cdot 10^{-8} \, T^2 + 9.88\cdot 10^{-5} \, T - 2.75\cdot 10^{-4},
\end{equation}
which described with a good accuracy, the evolution of the air thermal conductivity with temperature as shown in~\cite{Corpart2023}. The temperature is in Kelvin and the air thermal conductivity in ${\rm W \cdot m^{-1}\cdot K^{-1}}$.
\newpage

%%%%%%%%%%%%%%%%%%%%%
%
%%%%%%%%%%%%%%%%%%%%%
\section{Evaporation of a spherical drop}\label{sec:APPENDIX_spherical_drop}
In this appendix we detail the calculations of evaporation of a spherical drop first in purely diffusive regime then under forced convection.

\subsection{Diffusion-limited evaporation}\label{subsec:APPENDIX_sphere_diff}
\paragraph{Model}

We consider the mass transfer of the water vapor in the atmosphere surrounding a spherical drop of radius $R(t)$ and we assume that this process is limited by diffusion, which is valid in a quiescent atmosphere.
This is true for droplet radius significantly larger than the mean-free path of the vapor molecules, \textit{i.e.} $R$ larger than few micrometers~\cite{Fuchs1959}.
Over a timescale $R_0^2 / {\cal D}$, where $R_0$ is the initial radius, the transfer can be considered to occur in a stationary regime.
In practice, we can check that this timescale is short compared to the total evaporating time, such that the contribution of the starting non-stationary regime is negligible.

Thus, the concentration field $c$ is the solution of the Laplace equation $ \triangle c = 0$, which writes in spherical coordinates
\begin{equation}\label{eq:APPENDIX-mass_laplace}
       \frac{1}{r^2}\frac{\mathrm{d}}{\mathrm{d}r}\left(r^2\frac{\mathrm{d}c}{\mathrm{d}r}\right)  = 0.
\end{equation}
This equation is supplemented by two boundary conditions on the concentration, respectively at the liquid-vapor interface and far from the interface,
\begin{align}
    c(r = R) &= c_{\rm sat }(T_{\rm i}), \\
    c(r \to \infty) &= c_\infty,
\end{align}
where $T_{\rm i}$ is the temperature of the interface.
The relative humidity is defined as $\mathcal{R}_{\rm H} = p_\infty/P_{\rm sat}(T_\infty) \approx  c_\infty / c_{\rm sat}(T_\infty)$ in the ideal gas approximation, where $T_\infty$ is the air temperature far from the droplet.

By integrating equation~(\ref{eq:APPENDIX-mass_laplace}), the local evaporative flux given by Fick's law, $j = - {\cal D} \left. \frac{\mathrm{d}c}{\mathrm{d}r}\right|_{r = R}$, writes
\begin{equation}
    j = {\cal D} \frac{ c_{\rm sat }(T_{\rm i}) - c_\infty}{R}.
\end{equation}

The integration of the local flux over the evaporating surface gives $\Phi_{\rm ev} = \int j\, {\rm d} S = 4 \pi R \mathcal{D}(T_{\rm i}) ( c_{\rm sat }(T_{\rm i}) - c_\infty)$, which can be rewritten
\begin{equation}\label{eq:APPENDIX-sphere_diff_phi_ev}
\Phi_{\rm ev} = 4 \pi R {\cal D} c_{\rm sat}(T_\infty) \left(\frac{c_{\rm sat}(T_{\rm i})}{c_{\rm sat}(T_\infty)} - \mathcal{R}_{\rm H}\right).
\end{equation}
where ${\cal D}$ is in good approximation the vapor diffusion coefficient at $T = T_\infty$~\cite{Corpart2023}.

To compute the evaporation rate $\Phi_{\rm ev}$ the temperature of the liquid must be determined. To do so, we write in the next paragraph the heat transfer between the atmosphere and the drop.

As for the mass transfer, we consider a diffusion limited process in a stationary regime, for which, the air temperature field is a solution of the Laplace equation $ \triangle T = 0$ with the boundary conditions $T(r = R) = T_{\rm i}$ and $T(r \to \infty) = T_\infty$.
The steady-state assumption also implies that the temperature in the drop has reached its equilibrium value $T_{\rm i}$ and is uniform in the liquid.
This is validated if the timescale over which the heat diffuses through the liquid $R_0^2/\kappa_{\ell}$  with $\kappa_{\ell}$ the thermal conductivity of the liquid, is short compared to the evaporative time~\cite{Sobac2015}.
In practice, this is valid for the tested liquids evaporating under ambient conditions~\cite{Beard1971, Andreas1995, Sobac2015, Netz2020, Netz2020a}.

The integration of the Laplace equation leads to a total heat flux
\begin{equation}
    Q_{\rm h} = -4 \pi R \lambda_{\rm air} \Delta T ^\star,
\end{equation}
where $\Delta T ^\star = T_\infty - T_{\rm i}$.

The heat and mass fluxes are coupled through the enthalpy of vaporization $\Delta_{\rm vap} H$, $ \Delta_{\rm vap} H\,\Phi_{\rm ev} = - Q_{\rm h}$, which gives
\begin{equation}\label{eq:APPENDIX-Delta_T_vs_Delta_c_sphere_diff}
    \Delta T^\star = \chi\left(\frac{c_{\rm sat}(T_{\rm i})}{c_{\rm sat}(T_\infty)} - \mathcal{R}_{\rm H}\right),
\end{equation}
with $\chi = \frac{\Delta_{\rm vap} H {\cal D} c_{\rm sat}(T_\infty)}{ \lambda_{\rm air}}$ where $\Delta_{\rm vap} H$ is independent of the temperature (see Table~\ref{tab:APPENDIX-C_constantes_physiques_liq}), and the values of ${\cal D}, c_{\rm sat}(T_\infty) \text{ and } \lambda_{\rm air}$ are evaluated at $T = T_\infty$ by using the semi-empirical equations~(\ref{eq:APPENDIX-Fuller_equation}), (\ref{eq:APPENDIX-C_Antoine_equation}) and (\ref{eq:APPENDIX-lambda_air_vs_T_Andreas95}) respectively.

To solve equation~(\ref{eq:APPENDIX-Delta_T_vs_Delta_c_sphere_diff}) we introduce a quadratic approximation of $c_{\rm sat}(T)$, defined as
\begin{equation}\label{eq:APPENDIX-quadratic_c_sat_vs_T}
    c_\textrm{sat}(T) = c_\textrm{sat}(T_\infty)\left(1 + \alpha_1  (T_\infty - T) + \alpha_2  (T_\infty - T)^2 \right),
\end{equation}
where $\alpha_1$ and $\alpha_2$ are obtained by fitting the data from the literature (see Table~\ref{tab:APPENDIX-C_alpha_1_et_alpha_2}).
As discussed in~\cite{Corpart2023}, equation~(\ref{eq:APPENDIX-quadratic_c_sat_vs_T}) is an excellent approximation of Antoine's equation~(\ref{eq:APPENDIX-C_Antoine_equation}).

Combining equations~(\ref{eq:APPENDIX-Delta_T_vs_Delta_c_sphere_diff}) and (\ref{eq:APPENDIX-quadratic_c_sat_vs_T}), we get
\begin{equation} \label{eq:APPENDIX-eq_polynomiale_Delta_T}
    \chi \alpha_2 \Delta T^{\star 2} + \left(\chi \alpha_1 - 1 \right) \Delta T^\star + \chi \left(1 - \mathcal{R}_{\rm H} \right) = 0.
\end{equation}
Among the two roots admitted by equation (\ref{eq:APPENDIX-eq_polynomiale_Delta_T}), we keep the one for which $T_\infty - T_{\rm i}$ decreases as $ \mathcal{R}_{\rm H} $ increases, \textit{i.e.}

\begin{equation}\label{eq:APPENDIX-sphere_Delta_T_solution}
    \Delta T^\star = \frac{ 1 -  \chi \alpha_1  - \sqrt{\left(1 - \chi \alpha_1 \right)^2 - 4 \chi^2 \alpha_2 \left(1 - \mathcal{R}_{\rm H} \right)}}{2 \chi \alpha_2}.
\end{equation}

Inserting the above expression into equation~(\ref{eq:APPENDIX-sphere_diff_phi_ev}) we obtain the evaporation rate of the drop:
\begin{equation}
    \Phi_{\rm ev} = 4 \pi R {\cal D} c_{\rm sat}(T_\infty) \left[\alpha_2 \Delta T^{\star2}  + \alpha_1\Delta T^\star + 1 - \mathcal{R}_{\rm H} \right],
\end{equation}
with $\Delta T^\star$ given by equation~(\ref{eq:APPENDIX-sphere_Delta_T_solution}).

The droplet lifetime is obtained from the conservation of the drop volume $\Omega = \frac{4}{3} \pi R^3$,
\begin{equation}\label{eq:APPENDIX-mass_conservation}
    \Phi_{\rm ev} = - \rho \frac{{\rm d} \Omega}{{\rm d} t},
\end{equation}
where $\rho$ is the liquid density.
After integration from $R(0)=R_0$ to $R(\tau_{\rm S})=0$, we have the dynamics of the droplet radius $R(t) = R_0 \sqrt{1 - t / \tau_{\rm S}}$, where the droplet lifetime is
\begin{equation}\label{eq:APPENDIX_tau_sphere_diff}
    \tau_{\rm S} = \frac{\rho R_0^2}{2 \mathcal{D}c_{\rm sat}(T_{\infty}) \left(\alpha_2 \Delta T^{\star2}   + \alpha_1\Delta T^\star + 1 - \mathcal{R}_{\rm H} \right)},
\end{equation}
with $\Delta T^\star$ given by equation~(\ref{eq:APPENDIX-sphere_Delta_T_solution}).
In the following we compare lifetimes of a spherical drop and a drop on a fiber in purely diffusive regime.
\paragraph{Comparison with experiments}

\begin{figure}[h!]
    \centering
    \includegraphics[width = .6\linewidth]{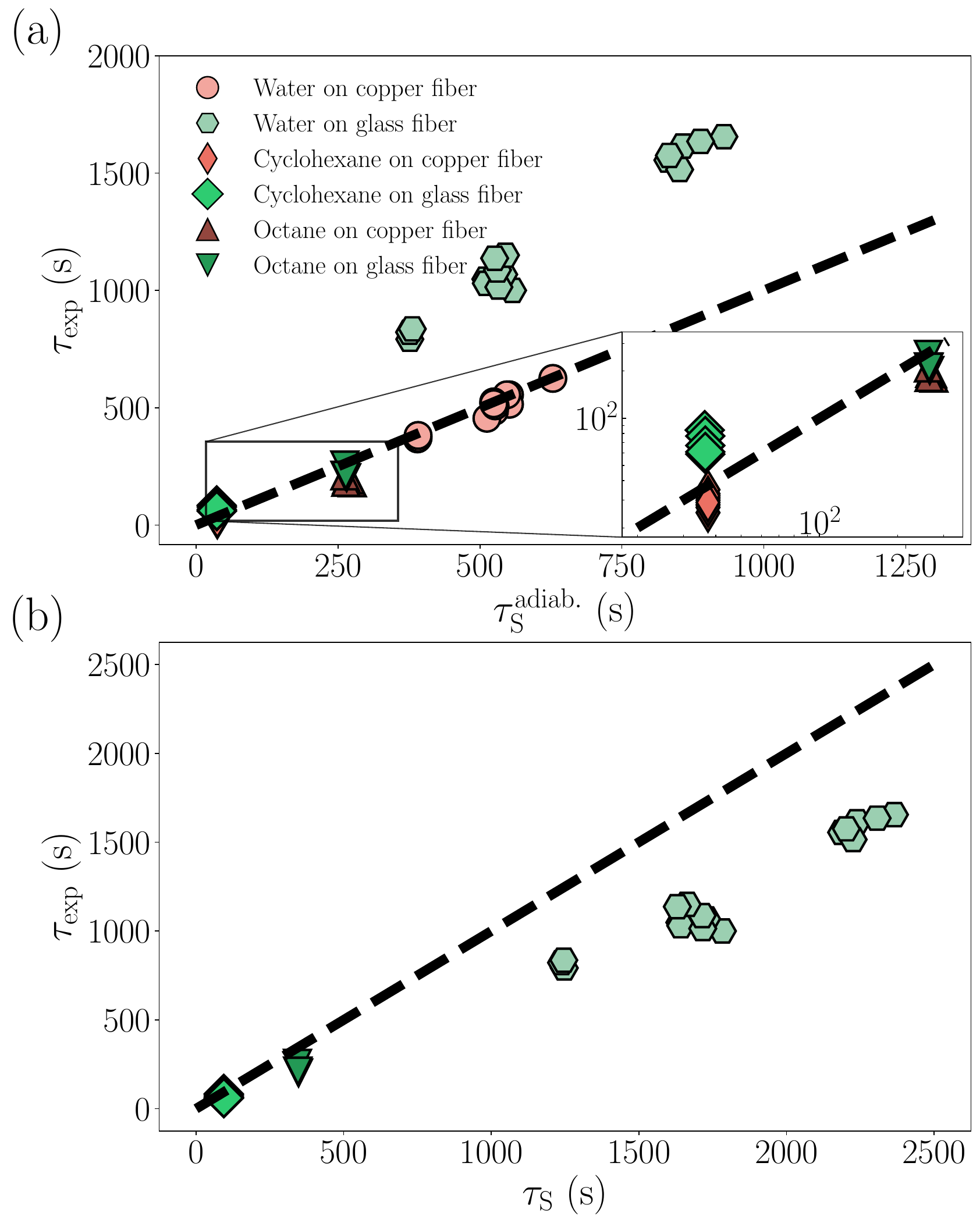}
    \caption{Comparison between lifetime of  an axisymmetric drop on a fiber and a spherical airborne drop evaporating in diffusive regime. (a) Measured lifetime of a drop on a fiber as a function of the lifetime of a spherical droplet (Eq.~(\ref{eq:APPENDIX_tau_sphere_diff})) evaporating adiabatically ($T_{\rm i} = T_\infty$) in the same conditions. (b) Measured lifetime of a drop on a glass fiber as a function of the lifetime of a spherical droplet (Eq.~(\ref{eq:APPENDIX_tau_sphere_diff})) evaporating and exchanging heat with air in the same conditions. The temperature of the liquid is given by equation~(\ref{eq:APPENDIX-sphere_Delta_T_solution}).}
    %\TODO{Caption + timelapse and raw data}}
    \label{fig:SM_tau_exp_vs_tau_sphere_DS}
\end{figure}
Figure~\ref{fig:SM_tau_exp_vs_tau_sphere_DS} shows a comparison between the measured lifetime of an axisymmetric drop on a fiber and the lifetime of a spherical drop evaporating under the same experimental conditions.

Figure~\ref{fig:SM_tau_exp_vs_tau_sphere_DS}(a) shows the experimental lifetimes plotted against the lifetime of a spherical drop evaporating adiabatically ($T_{\rm i} = T_\infty$) under the same experimental conditions (liquid, initial volume, temperature, relative humidity).
The dashed black curve in figure~\ref{fig:SM_tau_exp_vs_tau_sphere_DS}(a) represents equality between the axes. As explained in the main text, the adiabatic model describes very well the experimental results obtained for all the liquids tested placed on a copper wire, as well as the results obtained for octane placed on a glass fiber. On the other hand, there is a systematic discrepancy between the ideal case of a spherical drop evaporating adiabatically and what is measured for a drop of water or cyclohexane placed on a glass fiber.

Figure~\ref{fig:SM_tau_exp_vs_tau_sphere_DS}(b) displays the lifetime of different liquids evaporating on a glass fiber as a function of the lifetime of an airborne sphere composed of the same liquid, and cooling while evaporating under the same experimental conditions. The dotted curve represents axis equality. As mentioned in the main text, the presence of the fiber, even if it is made of an excellent thermal insulator, cannot be ignored in heat exchanges which explains that there is a systematic discrepancy between the experimental results obtained and the sphere model.

\subsection{Forced-convective evaporation}\label{subsec:APPENDIX_sphere_conv}
We now turn to evaporation of a spherical drop under forced convection.
As mentioned in the main text, Frössling~\cite{Frossling1938} showed that the evaporation rate of a spherical drop placed in a laminar flow can be written as :

\begin{equation}\label{eq:APPENDIX-Phi_ev_conv_sphere}
     \Phi_{\rm ev}^{\rm conv} =  \Phi_{\rm ev} f_{\rm ev},
\end{equation}
where $\Phi_{\rm ev}$ is the purely diffusive evaporation rate (Eq.~(\ref{eq:APPENDIX-sphere_diff_phi_ev})) and $f_{\rm ev}$ the Sherwood number or the ventilation coefficient:
\begin{equation}
    f_{\rm ev} = 1 + \beta_{\rm ev} \textrm{Re}^{1/2} \textrm{Sc}^{1/3}
\end{equation}

Analogously, the convective heat flux received by the drop from the atmosphere can be written as follows~\cite{Ranz1952a, Ranz1952}:
\begin{equation}\label{eq:APPENDIX-Q_h_conv_sphere}
     Q_{\rm h}^{\rm conv} =  Q_{\rm h}f_{\rm h}
\end{equation}
with $f_{\rm h}$ the Nusselt number or ventilation coefficient
\begin{equation}
    f_{\rm h} = 1 + \beta_{\rm h} \textrm{Re}^{1/2} \textrm{Pr}^{1/3},
\end{equation}
In the previous equations, $\mathrm{Re} = 2 R U_{\rm air}/\nu_{\rm air}$ is the Reynolds number characteristic of the flow, $\mathrm{Sc} = \nu_{\rm air}/\mathcal{D}$ is the Schmidt number, $\mathrm{Pr} = \nu_{\rm air}/{\alpha_{\rm air}}$ is the Prandtl number and $\beta_{\rm ev} \approx \beta_{\rm h} \approx 0.3$ is a constant. In gases, $\mathrm{Pr} \approx \mathrm{Sc} \approx 1$ so $f_{\rm ev} \approx f_{\rm h}$.

Writing the energy balance $\Delta_{\rm vap} H\Phi_{\rm ev}^{\rm conv} = - Q_{\rm h}^{\rm conv}$ we obtain the liquid temperature:

\begin{equation}\label{eq:APPENDIX-Delta_T_vs_Delta_c_sphere_CF}
    \Delta T^\star = \chi \frac{f_{\rm ev}}{f_{\rm h}} \left(\frac{c_{\rm sat}(T_{\rm i})}{c_{\rm sat}(T_\infty)} - \mathcal{R}_{\rm H}\right),
\end{equation}
which is virtually independent of air velocity and radius of the drop and approximately equal to the drop of temperature obtained in the purely diffusive regime (Eq.~(\ref{eq:APPENDIX-Delta_T_vs_Delta_c_sphere_diff})) whose solution is given by equation~(\ref{eq:APPENDIX-sphere_Delta_T_solution}).
Integrating the mass conservation, $\Phi_{\rm ev}^{\rm conv} = -\rho {\rm d} \Omega/{\rm d} t$ we obtain the lifetime of the drop evaporating in forced convection:

\begin{equation}
    \tau_{\rm S}^{\rm conv} = \frac{\rho}{{\cal D}c_{\rm sat}(T_\infty)\left(\alpha_2 \Delta T^{\star2}   + \alpha_1\Delta T^\star + 1 - \mathcal{R}_{\rm H} \right)} \, \int_{0}^{R_0} \frac{r \mathrm{d}r}{1 + Br^{1/2}},
\end{equation}
where $B= \beta_{\rm ev} \mathrm{Sc}^{1/3} \sqrt{2U_{\rm air}/\nu_{\rm air}}$ and $\Delta T^\star$ is given by equation~(\ref{eq:APPENDIX-sphere_Delta_T_solution}). The integration of the previous equation leads to the expression of the lifetime of the spherical drop of equation~(17) of the main text.

%%%%%%%%%%%%%%%%%%%%%
%
%%%%%%%%%%%%%%%%%%%%%

\section{Temperature of axisymmetric drop on a fiber}\label{sec:APPENDIX_temperature_drop_on_fiber}

\begin{figure}[h!]
    \centering
    \includegraphics[width = 1\linewidth]{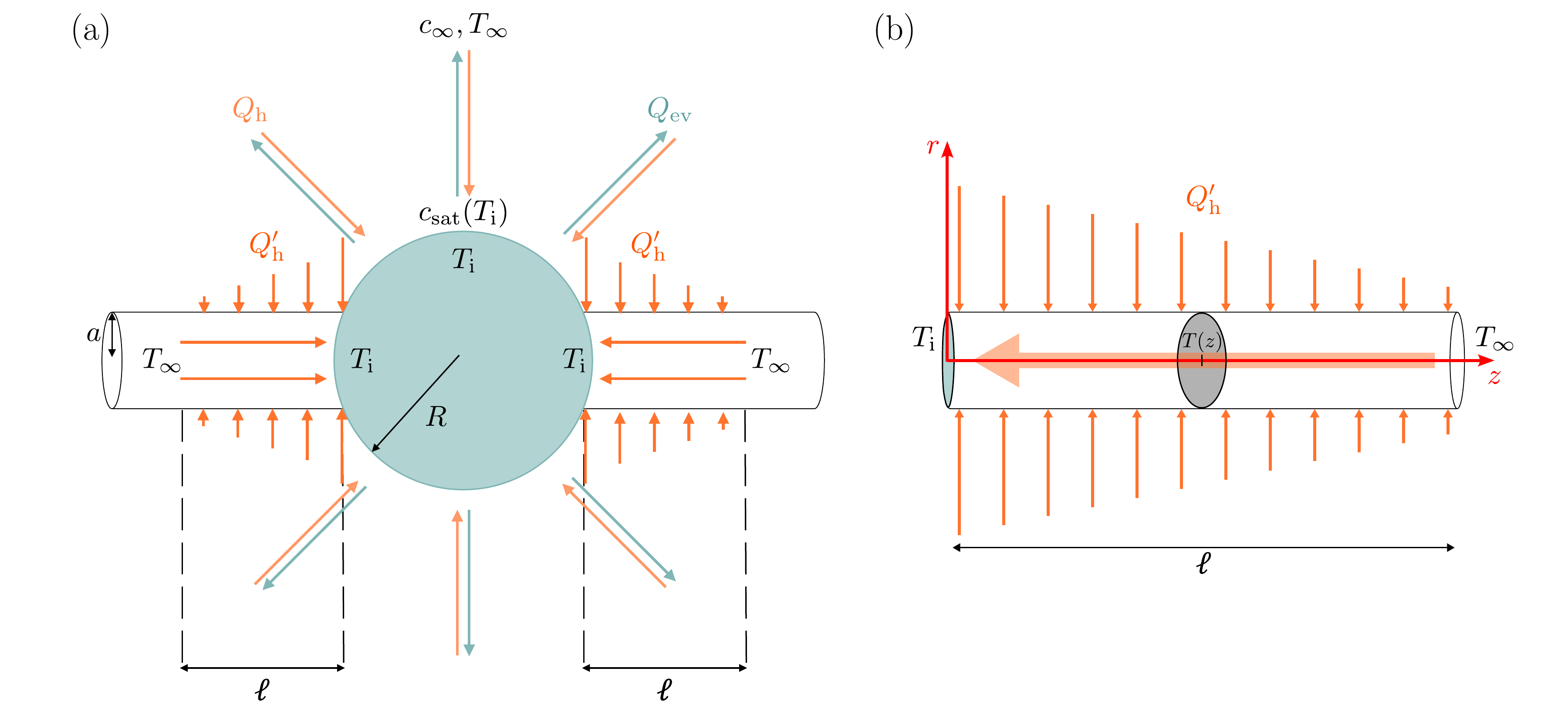}
    \caption{(a) Schematic representation of the heat fluxes exchanged between the drop on a fiber and its environment. (b) Coordinate system used to calculate the temperature in the fiber.}
    \label{fig:APPENDIX-scheme_heat_fluxes}
\end{figure}

To predict the temperature of an axisymmetric drop on a fiber we study the model system of a spherical drop pierced by a fiber show in figure~\ref{fig:APPENDIX-scheme_heat_fluxes}. 
We adopt the hypotheses listed in the main text of the articles, and here we detail the resolution of equation~(9) of the main text:
\begin{equation}\label{eq:APPENDIX-local_energy_balance_in_the_fiber}
    -2 \pi a \, \lambda_\textrm{air} \,  \left. \frac{\mathrm{d}T}{\mathrm{d}r} \right|_{r = a} = \pi a^2 \lambda_{\rm s}  \frac{\mathrm{d}^2T}{\mathrm{d}z^2}.
\end{equation}

The coordinate system is represented schematically in figure~\ref{fig:APPENDIX-scheme_heat_fluxes}(b).
To solve this differential equation we must estimate the temperature gradient in air (left-hand side of equation~(\ref{eq:APPENDIX-local_energy_balance_in_the_fiber})).

In a previous work~\cite{Corpart2022} we obtained analytically the surfacique evaporative flux of a liquid cylinder which is a problem analogous to the one that is considered here. We consider a cylinder of length $\ell \gg a$, shown in figure~\ref{fig:APPENDIX-scheme_heat_fluxes}(b), which receives a purely diffusive heat flux from the atmosphere with the following boundary conditions (i) the temperature at the surface of the dry fiber is equal to $T(z)$ and (ii) the temperature of the air far from the fiber is equal to $T_\infty$. By analogy between mass and heat transport, we can obtain the temperature gradient in air, evaluated at the surface of the fiber, in $r = a$, from equation~(4) of~\cite{Corpart2022}:

\begin{equation}\label{eq:APPENDIX-FIBRE_grad_T_dans_air}
    \left. \frac{\mathrm{d}T}{\mathrm{d}r}\right|_{r = a} = -\frac{\pi (T_\infty - T(z))}{2a\, f(\Tilde{\ell}/2)},
\end{equation}
where $\Tilde{\ell} = \ell/a$ and $f(x) =  2- 2 \gamma_e + \ln{2} + \frac{\pi}{2} \ln{(x)} \approx 10$, for $\ell \gg a$.
As shown in~\cite{Corpart2022}, the previous equation describes very well the results obtained numerically for the evaporation of a sleeve of finite length and large aspect ratio (slender cylinder). Equation~(\ref{eq:APPENDIX-FIBRE_grad_T_dans_air}) is therefore expected to be a good approximation of the temperature gradient in air in the situation considered here.
Inserting equation~(\ref{eq:APPENDIX-FIBRE_grad_T_dans_air}) into equation~(\ref{eq:APPENDIX-local_energy_balance_in_the_fiber}) and defining $\Theta(z) = T_\infty -T(z)$, we obtain the following differential equation:

\begin{equation}\label{eq:APPENDIX_FIBRE_eq_diff_T_de_z}
    \frac{\mathrm{d}^2\Theta}{\mathrm{d}z^2} = \frac{\Theta}{{\cal L}^2},
\end{equation}
whose solution is written 
\begin{equation}\label{eq:APPENDIX-FIBER_solution_T_z}
    T_\infty - T(z) = (T_\infty - T_{\rm i})\exp{\left(-\frac{z}{\cal L}\right)}.
\end{equation}
where the length ${\cal L}$ is given by:

\begin{equation}
    {\cal L} \approx  a\sqrt{\frac{10\lambda_{\rm s}}{\pi \lambda_{\rm air}}}.
\end{equation}

Inserting equation~(\ref{eq:APPENDIX-FIBER_solution_T_z}) into $Q{'}_h = - \pi a^2 \lambda_\textrm{s} \left. \overrightarrow{\nabla}T \right|   _{z = 0},$ (Eq~(8) of the main text), we get the heat flux exchanged between the fiber and the drop on one side:
\begin{equation}\label{eq:APPENDIX-Q_prime_h}
    Q{'}_\textrm{h}  \approx - \pi a \Delta T^\star  \sqrt{\frac{\pi \lambda_\textrm{air} \lambda_\textrm{s}}{10}}.
\end{equation}
This calculation gives similar results to what was done by Fuchs~\cite{Fuchs1959} for a spherical drop suspended at the tip of a fiber. To solve equation~(\ref{eq:APPENDIX-local_energy_balance_in_the_fiber}), he considers an infinite cylinder and solves the Laplace's equation for the temperature in air by introducing a cut-off length which must be large compare to all the other lengths of the problem. The obtained heat flux exchanged between the fiber and the drop is similar to the result of equation~(\ref{eq:APPENDIX-Q_prime_h}) for a cut-off length equals to 500 times the radius of the fiber.

Inserting equation~\ref{eq:APPENDIX-Q_prime_h} into the globale energie balance of the system, $\Delta_{\rm vap} H$, $ \Delta_{\rm vap} H\,\Phi_{\rm ev} = - (Q_{\rm h} + 2Q{'}_\textrm{h})$ gives the temperature of a drop on a fiber:

\begin{equation}\label{eq:APPENDIX-Delta_T_sphere_on_fiber}
        \Delta T^\star = \frac{ 1 + \Tilde{\mathcal{Q}}_{\rm fiber} -  \chi \alpha_1}{2 \chi \alpha_2} -  \frac{\sqrt{\left[1 +  \Tilde{\mathcal{Q}}_{\rm fiber} - \chi \alpha_1 \right]^2 - 4 \chi^2 \alpha_2 \left(1 - \mathcal{R}_{\rm H} \right)}}{2 \chi \alpha_2},
\end{equation}

From equation \ref{eq:APPENDIX-Delta_T_sphere_on_fiber}, asymptotic values for the temperature drop in the liquid can be obtained by comparing $\Tilde{\mathcal{Q}}_{\rm fiber}$ with $1 -\kappa \alpha_1 $.

For $\Tilde{\mathcal{Q}}_{\rm fiber} \ll 1 - \chi\alpha_1$ we fine the case of an airborne sphere:

\begin{equation}\label{eq:APPENDIX-fibre_asymptote_Delta_T_Q_f_petit}
    T_\infty - T_{\rm i} \approx \frac{ 1 -  \chi \alpha_1  - \sqrt{\left(1 - \chi \alpha_1 \right)^2 - 4 \chi^2 \alpha_2 \left(1 - \mathcal{R}_{\rm H} \right)}}{2 \chi \alpha_2}.
\end{equation}

% \paragraph{Cas où $Qfibre \gg  1 - \kappa\alpha_1 \approx 3$}
For $\Tilde{\mathcal{Q}}_{\rm fiber} \gg 1 - \chi\alpha_1$ we obtain :

% \begin{equation}
%     T_\infty - T_{\rm i} \approx \frac{\Qfibre \left( 1 - \sqrt{1 - \frac{4\kappa^2 \alpha_2 (1 - \mathcal{R}_{\rm H})}{\Qfibre^2}}\right)}{2 \kappa \alpha_2}.
% \end{equation}
% Or, pour l'eau, ${4\kappa^2\alpha_2} \approx 6 \ll \Qfibre^2$, dans le cas où $\Qfibre \ll 1 - \kappa \alpha_1 \approx 3$. On peut donc écrire le développement limité du terme
% $\sqrt{1 - {4\kappa^2 \alpha_2 (1 - \mathcal{R}_{\rm H})}/{\Qfibre^2}}$, et ainsi obtenir :

\begin{equation}\label{eq:ch4_fibre_asymptote_Delta_T_Q_f_grand}
    T_\infty - T_{\rm i}\approx \frac{\kappa}{\Tilde{\mathcal{Q}}_{\rm fiber}} (1 - \mathcal{R}_{\rm H}),
\end{equation}
 The two above equations are plotted, respectively in dash-dotted and dashed lines, in Figure~3(a) of the main text.
\bibliography{biblio}

\begin{thebibliography}{10}

\bibitem{Langmuir1918}
I.~Langmuir.
\newblock The evaporation of small spheres.
\newblock {\em Phys. Rev.}, 12(5):368--370, 1918.

\bibitem{Frossling1938}
N.~Frossling.
\newblock Uber die verdunstung fallernder tropfen.
\newblock {\em Gerlands Beitr. Geophys.}, 52:170--216, 1938.

\bibitem{Andreas1995}
E.~L. Andreas.
\newblock The temperature of evaporating sea spray droplets.
\newblock {\em J. Atmos. Sci.}, 52(7):852--862, 1995.

\bibitem{Corpart2024}
M.~Corpart, F.~Restagno, and F.~Boulogne.
\newblock Measuring relative humidity from evaporation with a wet-bulb thermometer: {The} psychrometer.
\newblock {\em American Journal of Physics}, 92(1):36--42, January 2024.

\bibitem{Sobac2015}
B.~Sobac, P.~Talbot, B.~Haut, A.~Rednikov, and P.~Colinet.
\newblock A comprehensive analysis of the evaporation of a liquid spherical drop.
\newblock {\em Journal of Colloid and Interface Science}, 438:306 -- 317, 2015.

\bibitem{Netz2020}
R.~R. Netz.
\newblock Mechanisms of airborne infection via evaporating and sedimenting droplets produced by speaking.
\newblock {\em J. Phys. Chem. B}, 124(33):7093--7101, August 2020.

\bibitem{Netz2020a}
R.~R. Netz and W.~A. Eaton.
\newblock Physics of virus transmission by speaking droplets.
\newblock {\em Proc. Natl. Acad. Sci.}, 117(41):25209--25211, 2020.

\bibitem{Erbil2012}
H.~Yildirim Erbil.
\newblock Evaporation of pure liquid sessile and spherical suspended drops: {A} review.
\newblock {\em Advances in Colloid and Interface Science}, 170(1-2):67--86, January 2012.

\bibitem{Cazabat2010}
A.-M. Cazabat and G.~Guéna.
\newblock Evaporation of macroscopic sessile droplets.
\newblock {\em Soft Matter}, 6(12):2591, 2010.

\bibitem{Larson2014}
R.~G. Larson.
\newblock Transport and deposition patterns in drying sessile droplets.
\newblock {\em AIChE J.}, 60(5):1538--1571, 2014.

\bibitem{Brutin2015}
D.~Brutin.
\newblock {\em Droplet wetting and evaporation: from pure to complex fluids}.
\newblock Elsevier/AP, Academic Press, Amsterdam Boston Heidelberg, 2015.
\newblock OCLC: 914244959.

\bibitem{Wilson2023}
S.~K. Wilson and H.-M. D’Ambrosio.
\newblock Evaporation of sessile droplets.
\newblock {\em Annu. Rev. Fluid Mech.}, 55(1):481--509, 2023.

\bibitem{Brutin2022}
D.~Brutin and K.~Sefiane.
\newblock {\em Drying of Complex Fluid Drops: Fundamentals and Applications}, volume~14.
\newblock Royal Society of Chemistry, 2022.

\bibitem{Picknett1977}
R.~G. Picknett and R.~Bexon.
\newblock The evaporation of sessile or pendant drops in still air.
\newblock {\em Journal of Colloid and Interface Science}, 61(2):336 -- 350, 1977.

\bibitem{Birdi1993}
K.S. Birdi and D.T. Vu.
\newblock Wettability and the evaporation rates of fluids from solid surfaces.
\newblock {\em Journal of Adhesion Science and Technology}, 7(6):485--493, January 1993.

\bibitem{Shanahan1994}
M.~E.~R. Shanahan and C.~Bourgès.
\newblock Effects of evaporation on contact angles on polymer surfaces.
\newblock {\em International Journal of Adhesion and Adhesives}, 14(3):201--205, July 1994.

\bibitem{Bourges-monnier1995}
C.~Bourges-Monnier and M.~E.~R. Shanahan.
\newblock Influence of {Evaporation} on {Contact} {Angle}.
\newblock {\em Langmuir}, 11(7):2820--2829, July 1995.

\bibitem{Rowan1995}
S.~M. Rowan, M.~I. Newton, and G.~McHale.
\newblock Evaporation of {Microdroplets} and the {Wetting} of {Solid} {Surfaces}.
\newblock {\em J. Phys. Chem.}, 99(35):13268--13271, August 1995.

\bibitem{Mchale1998}
G.~McHale, S.~M. Rowan, M.~I. Newton, and M.~K. Banerjee.
\newblock Evaporation and the {Wetting} of a {Low}-{Energy} {Solid} {Surface}.
\newblock {\em J. Phys. Chem. B}, 102(11):1964--1967, March 1998.

\bibitem{Erbil2002}
H.~Yildirim Erbil, G.~McHale, and M.~I. Newton.
\newblock Drop {Evaporation} on {Solid} {Surfaces}: {Constant} {Contact} {Angle} {Mode}.
\newblock {\em Langmuir}, 18(7):2636--2641, April 2002.

\bibitem{Soolaman2005}
D.~M. Soolaman and H.-Z. Yu.
\newblock Water {Microdroplets} on {Molecularly} {Tailored} {Surfaces}: {Correlation} between {Wetting} {Hysteresis} and {Evaporation} {Mode} {Switching}.
\newblock {\em J. Phys. Chem. B}, 109(38):17967--17973, September 2005.

\bibitem{Stauber2014}
Jutta~M Stauber, Stephan~K Wilson, Brian~R Duffy, and Khellil Sefiane.
\newblock On the lifetimes of evaporating droplets.
\newblock {\em Journal of Fluid Mechanics}, 744:R2, 2014.

\bibitem{Stauber2015}
J.~M. Stauber, S.~K. Wilson, B.~R. Duffy, and K.~Sefiane.
\newblock On the lifetimes of evaporating droplets with related initial and receding contact angles.
\newblock {\em Physics of Fluids}, 27(12):122101, December 2015.

\bibitem{David2007}
S.~David, K.~Sefiane, and L.~Tadrist.
\newblock Experimental investigation of the effect of thermal properties of the substrate in the wetting and evaporation of sessile drops.
\newblock {\em Colloids and Surfaces A: Physicochemical and Engineering Aspects}, 298(1):108--114, April 2007.

\bibitem{Dunn2009}
G.~J. Dunn, S.~K. Wilson, B.~R. Duffy, S.~David, and K.~Sefiane.
\newblock The strong influence of substrate conductivity on droplet evaporation.
\newblock {\em J. Fluid Mech.}, 623:329--351, 2009.

\bibitem{Sefiane2011}
K.~Sefiane and R.~Bennacer.
\newblock An expression for droplet evaporation incorporating thermal effects.
\newblock {\em Journal of Fluid Mechanics}, 667:260--271, January 2011.

\bibitem{Lopes2013}
M.~C. Lopes, E.~Bonaccurso, T.~Gambaryan-Roisman, and P.~Stephan.
\newblock Influence of the substrate thermal properties on sessile droplet evaporation: {Effect} of transient heat transport.
\newblock {\em Colloids and Surfaces A: Physicochemical and Engineering Aspects}, 432:64--70, September 2013.

\bibitem{Schofield2021}
Feargus~GH Schofield, David Pritchard, Stephen~K Wilson, and Khellil Sefiane.
\newblock The lifetimes of evaporating sessile droplets of water can be strongly influenced by thermal effects.
\newblock {\em Fluids}, 6(4):141, 2021.

\bibitem{Dehaeck2014}
S.~Dehaeck, A.~Rednikov, and P.~Colinet.
\newblock Vapor-based interferometric measurement of local evaporation rate and interfacial temperature of evaporating droplets.
\newblock {\em Langmuir}, 30(8):2002--2008, 2014.

\bibitem{Hu2005}
H.~Hu and R.~G. Larson.
\newblock Analysis of the {Effects} of {Marangoni} {Stresses} on the {Microflow} in an {Evaporating} {Sessile} {Droplet}.
\newblock {\em Langmuir}, 21(9):3972--3980, April 2005.

\bibitem{Ristenpart2007}
W.~D. Ristenpart, P.~G. Kim, C.~Domingues, J.~Wan, and H.~A. Stone.
\newblock Influence of substrate conductivity on circulation reversal in evaporating drops.
\newblock {\em Phys. Rev. Lett.}, 99:234502, Dec 2007.

\bibitem{Xu2010}
X.~Xu, J.~Luo, and D.~Guo.
\newblock Criterion for {Reversal} of {Thermal} {Marangoni} {Flow} in {Drying} {Drops}.
\newblock {\em Langmuir}, 26(3):1918--1922, February 2010.

\bibitem{Girard2008}
F.~Girard, M.~Antoni, and K.~Sefiane.
\newblock On the {Effect} of {Marangoni} {Flow} on {Evaporation} {Rates} of {Heated} {Water} {Drops}.
\newblock {\em Langmuir}, 24(17):9207--9210, September 2008.

\bibitem{Semenov2010}
S.~Semenov, V.~M. Starov, R.~G. Rubio, and M.~G. Velarde.
\newblock Instantaneous distribution of fluxes in the course of evaporation of sessile liquid droplets: {Computer} simulations.
\newblock {\em Colloids and Surfaces A: Physicochemical and Engineering Aspects}, 372(1):127--134, December 2010.

\bibitem{Yang2022}
X.~Yang, M.~Wu, M.~Doi, and X.~Man.
\newblock Evaporation {Dynamics} of {Sessile} {Droplets}: {The} {Intricate} {Coupling} of {Capillary}, {Evaporation}, and {Marangoni} {Flow}.
\newblock {\em Langmuir}, 38(16):4887--4893, April 2022.

\bibitem{Pan2006}
N.~Pan and P.~Gibson.
\newblock {\em Thermal and Moisture Transport in Fibrous Materials}.
\newblock Woodhead Publishing Series in Textiles. Elsevier Science, 2006.

\bibitem{Sutter2010}
B.~Sutter, D.~Bémer, J.-C. Appert-Collin, D.~Thomas, and N.~Midoux.
\newblock Evaporation of {Liquid} {Semi}-{Volatile} {Aerosols} {Collected} on {Fibrous} {Filters}.
\newblock {\em Aerosol Science and Technology}, 44(5):395--404, March 2010.

\bibitem{Duprat2022}
C.~Duprat.
\newblock Moisture in textiles.
\newblock {\em Annu. Rev. Fluid Mech.}, 54(1):443--467, 2022.

\bibitem{Carroll1976}
B.J. Carroll.
\newblock The accurate measurement of contact angle, phase contact areas, drop volume, and laplace excess pressure in drop-on-fiber systems.
\newblock {\em Journal of Colloid and Interface Science}, 57(3):488 -- 495, 1976.

\bibitem{Chou2011}
T.-H. Chou, S.-J. Hong, Y.-E. Liang, H.-K. Tsao, and Y.-J. Sheng.
\newblock Equilibrium phase diagram of drop-on-fiber: Coexistent states and gravity effect.
\newblock {\em Langmuir}, 27(7):3685--3692, 2011.

\bibitem{Mchale2002}
G.~McHale and M.I. Newton.
\newblock Global geometry and the equilibrium shapes of liquid drops on fibers.
\newblock {\em Colloids and Surfaces A: Physicochemical and Engineering Aspects}, 206(1-3):79--86, 2002.

\bibitem{Brochard1991}
Fran{\c{c}}oise Brochard-Wyart, Jean~Marc Di~Meglio, David Qu{\'e}re, and Pierre~Gilles De~Gennes.
\newblock Spreading of nonvolatile liquids in a continuum picture.
\newblock {\em Langmuir}, 7(2):335--338, 1991.

\bibitem{Corpart2022}
M.~Corpart, J.~Dervaux, C.~Poulard, F.~Restagno, and F.~Boulogne.
\newblock Evaporation of liquid coating a fiber.
\newblock {\em Europhysics Letters}, 139(4):43001, August 2022.

\bibitem{Duprat2013}
C.~Duprat, A.~D. Bick, P.~B. Warren, and H.~A. Stone.
\newblock Evaporation of drops on two parallel fibers: Influence of the liquid morphology and fiber elasticity.
\newblock {\em Langmuir}, 29(25):7857--7863, 2013.

\bibitem{Boulogne2015a}
F.~Boulogne, A.~Sauret, B.~Soh, E.~Dressaire, and H.~A. Stone.
\newblock Mechanical tuning of the evaporation rate of liquid on crossed fibers.
\newblock {\em Langmuir}, 31(10):3094--3100, 2015.

\bibitem{Chauveau2019}
C.~Chauveau, M.~Birouk, F.~Halter, and I.~Gökalp.
\newblock An analysis of the droplet support fiber effect on the evaporation process.
\newblock {\em International Journal of Heat and Mass Transfer}, 128:885--891, January 2019.

\bibitem{George2017}
O.~A. George, J.~Xiao, C.~S. Rodrigo, R.~Mercadé-Prieto, J.~Sempere, and X.~D. Chen.
\newblock Detailed numerical analysis of evaporation of a micrometer water droplet suspended on a glass filament.
\newblock {\em Chemical Engineering Science}, 165:33--47, June 2017.

\bibitem{Ghata2014}
N.~Ghata and B.~D. Shaw.
\newblock Computational modeling of the effects of support fibers on evaporation of fiber-supported droplets in reduced gravity.
\newblock {\em International Journal of Heat and Mass Transfer}, 77:22--36, October 2014.

\bibitem{Shih1995}
A.~T. Shih and C.~M. Megaridis.
\newblock Suspended droplet evaporation modeling in a laminar convective environment.
\newblock {\em Combustion and Flame}, 102(3):256--270, August 1995.

\bibitem{Shringi2013}
D.~Shringi, H.A. Dwyer, and B.D. Shaw.
\newblock Influences of support fibers on vaporizing fuel droplets.
\newblock {\em Computers \& Fluids}, 77:66--75, April 2013.

\bibitem{Yang2001}
J.-R. Yang and S.-C. Wong.
\newblock On the discrepancies between theoretical and experimental results for microgravity droplet evaporation.
\newblock {\em International Journal of Heat and Mass Transfer}, 44(23):4433--4443, December 2001.

\bibitem{Yang2002}
J.-R. Yang and S.-C. Wong.
\newblock An experimental and theoretical study of the effects of heat conduction through the support fiber on the evaporation of a droplet in a weakly convective flow.
\newblock {\em International Journal of Heat and Mass Transfer}, 45(23):4589--4598, November 2002.

\bibitem{Radhakrishnan2019}
S.~Radhakrishnan, N.~Srivathsan, T.N.C. Anand, and Shamit Bakshi.
\newblock Influence of the suspender in evaporating pendant droplets.
\newblock {\em International Journal of Thermal Sciences}, 140:368--376, June 2019.

\bibitem{Fuchs1959}
N.~A. Fuchs.
\newblock {\em Evaporation and {Droplet} {Growth} in {Gaseous} {Media}}.
\newblock Pergamon Press, London, 1959.

\bibitem{Bintein2019}
P.-B. Bintein, H.~Bense, C.~Clanet, and D.~Quéré.
\newblock Self-propelling droplets on fibres subject to a crosswind.
\newblock {\em Nature Physics}, 15(10):1027--1032, October 2019.

\bibitem{Wilson2023a}
J.~L. Wilson, A.~A. Pahlavan, M.~A. Erinin, C.~Duprat, L.~Deike, and H.~A. Stone.
\newblock Aerodynamic interactions of drops on parallel fibres.
\newblock {\em Nature Physics}, 2023.

\bibitem{Corpart2023}
M.~Corpart, F.~Restagno, and F.~Boulogne.
\newblock Analytical prediction of the temperature and the lifetime of an evaporating spherical droplet.
\newblock {\em Colloids and Surfaces A: Physicochemical and Engineering Aspects}, page 132059, 2023.

\bibitem{Beard1971}
K.~V. Beard and H.~R. Pruppacher.
\newblock A wind tunnel investigation of the rate of evaporation of small water drops falling at terminal velocity in air.
\newblock {\em J. Atmos. Sci.}, 28(8):1455 -- 1464, 1971.

\bibitem{Worthington1885}
A.~M. Worthington.
\newblock {IV}. {Note} on a point in the theory of pendent drops.
\newblock {\em The London, Edinburgh, and Dublin Philosophical Magazine and Journal of Science}, 19(116):46--48, 1885.
\newblock Publisher: Taylor \& Francis.

\bibitem{Jones2001}
E.~Jones, T.~Oliphant, P.~Peterson, et~al.
\newblock {SciPy}: Open source scientific tools for {Python}, 2001--.

\bibitem{Ranz1952a}
W.~E. Ranz and W.R. Marshall.
\newblock Evaporation from drops: Part {II}.
\newblock {\em Chemical Engineering Progress}, 48(4):173--181, 1952.

\bibitem{Ranz1952}
W.~E. Ranz and W.R. Marshall.
\newblock Evaporation from drops: Part {I}.
\newblock {\em Chemical Engineering Progress}, 48(3):141--146, 1952.

\bibitem{Bintein2015}
P.-B. Bintein.
\newblock {\em Dynamiques de gouttes funambules : applications à la fabrication de laine de verre}.
\newblock phdthesis, Université Pierre et Marie Curie - Paris VI, January 2015.

\bibitem{Gelderblom2022}
H.~Gelderblom, C.~Diddens, and A~Marin.
\newblock Evaporation-driven liquid flow in sessile droplets.
\newblock {\em Soft Matter}, 18:8535--8553, 2022.

\bibitem{Hu2006}
H.~Hu and R.~G. Larson.
\newblock Marangoni {Effect} {Reverses} {Coffee}-{Ring} {Depositions}.
\newblock {\em J. Phys. Chem. B}, 110(14):7090--7094, April 2006.

\bibitem{Corpart2023JFM}
M.~Corpart, F.~Restagno, and F.~Boulogne.
\newblock Coffee stain effect on a fibre from axisymmetric droplets.
\newblock {\em J. Fluid Mech.}, 957:A24, February 2023.

\bibitem{Gupta2021}
Ankur Gupta, Andrew~R Konicek, Mark~A King, Azmaine Iqtidar, Mohsen~S Yeganeh, and Howard~A Stone.
\newblock Effect of gravity on the shape of a droplet on a fiber: Nearly axisymmetric profiles with experimental validation.
\newblock {\em Physical Review Fluids}, 6(6):063602, 2021.

\bibitem{Lide2008}
D.R. Lide, editor.
\newblock {\em CRC Handbook of Chemistry and Physics}.
\newblock CRC Press/Taylor and Francis, 89th edition edition, 2008.

\bibitem{Fuller1969}
E.N. Fuller, K.~Ensley, and J.C. Giddings.
\newblock Diffusion of halogenated hydrocarbons in helium. the effect of structure on collision cross sections.
\newblock {\em The Journal of Physical Chemistry}, 73(11):3679--3685, 1969.

\bibitem{Fuller1966}
E.~N. Fuller, P.~D. Schettler, and J.~C. Giddings.
\newblock New method for prediction of binary gas-phase diffusion coefficients.
\newblock {\em Industrial \& Engineering Chemistry}, 58(5):18--27, 1966.

\bibitem{Jakli1978}
G.~Jakli, P.~Tzias, and W.~A. Van~Hook.
\newblock Vapor pressure isotope effects in the benzene (b)--cyclohexane (c) system from 5 to 80° ci the pure liquids b-d 0, b-d 1, ortho-, meta-, and para-b-d 2, b-d 6, c-d 0, and c-d 12. ii. excess free energies and isotope effects on excess free energies in the solutions b-h 6/b-d 6, c-h 12/c-d 12, b-h 6/c-h 12, b-d 6/c-h 12, and b-h 6/c-d 12.
\newblock {\em The Journal of Chemical Physics}, 68(7):3177--3190, 1978.

\bibitem{Cook1958}
M.~W. Cook.
\newblock Vapor {Pressure} {Apparatus}.
\newblock {\em Review of Scientific Instruments}, 29(5):399--400, May 1958.

\bibitem{Young1928}
S.~Young.
\newblock On the boiling points of the normal paraffins at different pressures.
\newblock {\em Proceedings of the Royal Irish Academy. Section B: Biological, Geological, and Chemical Science}, 38:65--92, 1928.

\bibitem{Linder1931}
E.~G. Linder.
\newblock Vapor {Pressures} of {Some} {Hydrocarbons}.
\newblock {\em The Journal of Physical Chemistry}, 35(2):531--535, February 1931.

\bibitem{Young1900}
S.~Young.
\newblock {CIV}.—{Vapour} pressures, specific volumes, and critical constants of normal octane.
\newblock {\em Journal of the Chemical Society, Transactions}, 77(0):1145--1151, January 1900.

\bibitem{Dejoz1996}
A.~Dejoz, V.~González-Alfaro, P.~J. Miguel, and M.~I. Vázquez.
\newblock Isobaric {Vapor}--{Liquid} {Equilibria} for {Binary} {Systems} {Composed} of {Octane}, {Decane}, and {Dodecane} at 20 {kPa}.
\newblock {\em Journal of Chemical \& Engineering Data}, 41(1):93--96, January 1996.

\bibitem{Ewing2000}
M.B. Ewing and J.C. Sanchez~Ochoa.
\newblock The vapour pressure of cyclohexane over the whole fluid range determined using comparative ebulliometry.
\newblock {\em The Journal of Chemical Thermodynamics}, 32(9):1157--1167, September 2000.

\bibitem{Weiguo1990}
S.~Weiguo, A.~X. Qin, P.~J. McElroy, and A.~G. Williamson.
\newblock ({Vapour} + liquid) equilibria of (n-hexane + n-hexadecane), (n-hexane + n-octane), and (n-octane + n-hexadecane).
\newblock {\em The Journal of Chemical Thermodynamics}, 22(9):905--914, September 1990.

\bibitem{Carruth1973}
G.~F. Carruth and R.~Kobayashi.
\newblock Vapor pressure of normal paraffins ethane through n-decane from their triple points to about 10 mm mercury.
\newblock {\em Journal of Chemical \& Engineering Data}, 18(2):115--126, April 1973.

\bibitem{Bell1968}
T.~N. Bell, E.~L. Cussler, K.~R. Harris, C.~N. Pepela, and P.~J. Dunlop.
\newblock An apparatus for degassing liquids by vacuum sublimation.
\newblock {\em The Journal of Physical Chemistry}, 72(13):4693--4695, December 1968.

\bibitem{Cruickshank1967}
A.~J.~B. Cruickshank and A.~J.~B. Cutler.
\newblock Vapor pressure of cyclohexane, 25 to 75.degree.
\newblock {\em Journal of Chemical \& Engineering Data}, 12(3):326--329, July 1967.

\bibitem{Ksiazczak1991}
A.~Ksiazczak and J.~J. Kosinski.
\newblock Vapor pressure of solutions of polar aromatic compounds in cyclohexane at 298.15 and 323.15 {K}.
\newblock {\em Journal of Chemical \& Engineering Data}, 36(4):351--354, October 1991.

\bibitem{Washburn1935}
E.~R. Washburn and B.~H. Handorf.
\newblock The {Vapor} {Pressure} of {Binary} {Solutions} of {Ethyl} {Alcohol} and {Cyclohexane} at 25°.
\newblock {\em Journal of the American Chemical Society}, 57(3):441--443, March 1935.

\bibitem{Carmona2000}
F.~J. Carmona, J.~A. González, I.~García de~la Fuente, J.~C. Cobos, V.~R. Bhethanabotla, and S.~W. Campbell.
\newblock Thermodynamic {Properties} of \textit{n} -{Alkoxyethanols} + {Organic} {Solvent} {Mixtures}. {XI}. {Total} {Vapor} {Pressure} {Measurements} for \textit{n} -{Hexane}, {Cyclohexane} or \textit{n} -{Heptane} + 2-{Ethoxyethanol} at 303.15 and 323.15 {K}.
\newblock {\em Journal of Chemical \& Engineering Data}, 45(4):699--703, July 2000.

\end{thebibliography}

\bibliographystyle{unsrt}

\end{document}